\documentclass[times]{myarX}

\usepackage{moreverb}

\newcommand\BibTeX{{\rmfamily B\kern-.05em \textsc{i\kern-.025em b}\kern-.08em
T\kern-.1667em\lower.7ex\hbox{E}\kern-.125emX}}

\usepackage{siunitx}
\usepackage{graphicx}
\usepackage[labelformat=simple]{subcaption}

\usepackage{setspace}
\usepackage[framemethod=TikZ]{mdframed}
\usepackage{todonotes}
\usepackage[square,numbers,sort&compress]{natbib}

\renewcommand{\vec}[1]{\boldsymbol{#1}}

\renewcommand{\d}{\mbox{d}}

\DeclareSIUnit\Pascal{Pa}
\DeclareSIUnit\mmMercury{mmHg}

\begin{document}

\title{Fluid-structure interaction modelling and stabilisation of a patient-specific arteriovenous access fistula}

\author{W.~P.~Guess\affil{1}\textsuperscript{,}\affil{2}\textsuperscript{,}\footnotemark[2], B.~D. Reddy\affil{1}\textsuperscript{,}\affil{3}, A.~McBride\affil{1}\textsuperscript{,}\affil{4}, B.~Spottiswoode\affil{5}\textsuperscript{,}\affil{6}, J.~Downs\affil{7}, T.~Franz\affil{5}\textsuperscript{,}\affil{8}}

\address{\centering\affilnum{1} Centre for Research in Computational and Applied Mechanics (CERECAM), University of Cape Town, 7701 Rondebosch, South Africa\\
\affilnum{2} Department of Mechanical Engineering, Faculty of Engineering and the Built Environment, University of Cape Town, 7701 Rondebosch, South Africa\\ 
\affilnum{3} Department of Mathematics and Applied Mathematics, University of Cape Town, 7701 Rondebosch, South Africa\\
\affilnum{4}\ Infrastructure and Environment, School of Engineering, The University of Glasgow, Glasgow G12 8QQ\\
\affilnum{5}\ Division of Biomedical Engineering, Department of Human Biology, Faculty of Health Sciences, University of Cape Town, 7935 Observatory, South Africa\\
\affilnum{6}\ Siemens Medical Solutions USA, Inc., 40 Liberty Boulevard, Malvern, PA 19355, USA\\
\affilnum{7}\ Division of General Surgery, Department of Surgery, Groote Schuur Hospital, 7925 Observatory, South Africa\\
\affilnum{8}\ Bioengineering Science Research Group, Engineering Sciences, Faculty of Engineering and the Environment, University of Southampton, Southampton, UK
}

\begin{abstract}
	A patient-specific fluid-structure interaction (FSI) model of a phase-contrast magnetic resonance angiography (PC-MRA) imaged arteriovenous fistula is presented. The numerical model is developed and simulated using a commercial multiphysics simulation package where a semi-implicit FSI coupling scheme combines a finite volume method blood flow model and a finite element method vessel wall model. A pulsatile mass-flow boundary condition is prescribed at the artery inlet of the model, and a three-element Windkessel model at the artery and vein outlets. The FSI model is freely available for analysis and extension. This work shows the effectiveness of combining a number of stabilisation techniques to simultaneously overcome the added-mass effect and optimise the efficiency of the overall model. The PC-MRA data, fluid model, and FSI model results show almost identical flow features in the fistula; this applies in particular to a flow recirculation region in the vein that could potentially lead to fistula failure.
\end{abstract}


\maketitle

\footnotetext[2]{winston.guess@alumni.uct.ac.za}

\section{Introduction}
	Haemodialysis is a form of renal replacement therapy (RRT) that saves and extends the lives of millions patients worldwide \citep{Vazquez2009,Liyanage2015,FMC2014}. The morbidity, mortality rate and quality of life of these patients are however severely affected in part by the consequent complications and interventions associated with vascular access via a surgically created access point to the bloodstream \citep{schwab1999vascular,kdoqi2006}. These complications and interventions are also a large contributor to the high cost of caring for dialysis patients and curtail the growth of haemodialysis, especially in developing countries where it is most needed \citep{Nahas2005}.
	
	Renal replacement therapy is required when a patient has end stage renal disease (ESRD)and no longer has any endogenous renal function. A global market survey by Fresenius Medical Care estimated that by the end of 2013 there were more than 3 million ESRD patients worldwide \citep{FMC2014}. With the growth rate of the renal replacement population far exceeding the world population growth rate, this number is expected to rise to 5 million by the year 2030 \citep{FMC2014,Liyanage2015}. Estimates show that the number of patients currently needing RRT worldwide is already between 4.9 and 9.7 million \citep{Liyanage2015}. There is thus a great need to increase access to dialysis and reduce the RRT gap in developing countries \citep{Nahas2005,Barsoum2006}. Although the situation is less dire in developed countries, ESRD remains a major cost for health-care systems while the substantial growth in the prevalent dialysis population is expected to continue. 
	
	The treatment of choice for renal failure is the kidney transplant, providing patients with a much better quality of life than those on dialysis. Few patients are however able to receive a donor kidney, and for the few who do have access to transplantation, waiting lists are long. Haemodialysis is used in more than 70\% of the ESRD population. This proportion is expected to increase while the use of peritoneal dialysis, another form of dialysis, is expected to decrease in developed countries \citep{FMC2014, jain2012global}. In haemodialysis treatments blood is pumped along an extracorporeal circuit and through a dialyzer, a filter that acts as an artificial kidney, before being returned to the bloodstream. Patients are typically required to be connected to the dialysis machine for four hours, three times a week. A surgically placed vascular access provides access the bloodstream and generates the high flow rates required for efficient haemodialysis. The arteriovenous (AV) fistula, AV graft and central venous catheter are the three primary forms of vascular access. The high complication and failure rate of fistulas and grafts are usually as a result of the increased flow rate and the pulsatile nature of the blood flow generated by the access through its low resistance pathway, the vein. Ventral venous catheters (temporary or permanent) are prone to mechanical complications, infections, and thrombotic episodes and are associated with the highest rate of complications morbidity, and mortality of all access methods. They are therefore regarded as only a temporising measure or when no other access option is viable \citep{mcgee2003preventing,windus1993permanent, schwab1997nkf}. 
	
	Fistulas are formed by suturing a vein directly to an artery in the upper arm or forearm. The vein and anastomosis, the surgical connection between the artery and vein, then take between three to six months to mature for cannulation. Grafts use a length of prosthetic tubing tubing to connect the artery and vein; this allows for access in many positions across the body. A graft usually requires two to three weeks of healing after surgery before being ready for cannulation and dialysis while certain newer grafts take only a matter of hours \citep{Vazquez2009, hazinedaroglu2004immediate}. Although fistulas have a greater likelihood of maturation failure, especially in older patients and those with diabetes, once mature their patency rates increase drastically. They are likely to have a far longer life span than other access methods and in some cases can last for over ten years \citep{schwab1999vascular,Kharboutly2007}. The average life of a graft is two to three years, while catheters on average have a short life span of only twelve months at best \citep{kdoqi2006}.
	
	Numerous studies have found that fistulas have overall the best long term patency rates and the lowest required intervention rate \citep{Kumwenda2015, wasse2007association, allon2007current, ravani2007clinical}. The fistula is therefore considered the first choice of access and was found to be used in 60\% of the dialysis population in 2011 \citep{kdoqi2006,fan1992vascular,windus1993permanent, harland1994placement,munda1983}. In situations where patients cannot wait out a long maturation period, or have veins and arteries unsuitable for fistula access, a graft is the best choice of access, while catheters are used for immediate access, and as a bridge to other forms of access.
	
	The vein walls in AV access experience abnormally high and oscillatory blood flow rates, pressures, and shear stresses due to shunting and need to adapt successfully to remain patent. The increased flow and shear stresses lead to outward wall remodelling while the increase in pressure leads to an increase in tensile stress and in turn to vessel wall thickening. Maturation failure, from early neointimal hyperplasia and insufficient or excessive vein adaptation, is the most common fistula dysfunction \citep{lee2015new}. Numerous studies have also found that neointimal hyperplasia is responsible for the progression of venous outflow stenosis and eventual thrombosis, and is the leading cause of access failure especially in grafts \citep{safa1996detection, roy2001venous, lee2009advances}.
	
	Many approaches are being taken to better understand and find the haemodynamic factors that promote neointimal hyperplasia development in haemodialysis patients. Although a number of \textit{in-vivo} blood flow measuring techniques are available, they do not have a resolution capable of accurately determining the complex flow in access vessels. Recently much work has been done to numerically model the behaviour of blood flow in vascular access and other cardiovascular vessels. Computational fluid dynamics (CFD) simulations of patient-specific fistula have been able to approximate the haemodynamics at a high resolution and determine with relatively high accuracy, quantities such as pressure and wall shear stress (WSS) \citep{EneIordache2001, Kharboutly2007, niemann2011computational}. These studies have also shown that abnormal flow recirculation and WSS conditions exist in fistulas in regions that correlate with neointimal development.
	
	CFD models however do not capture the deformation of the vessel wall, decreasing the accuracy of the final results and possibly obscuring valuable flow information that might occur due to the interaction with highly-deformable vessel tissue. Numerous studies have incorporated vessel wall mechanics in FSI simulations to simulate more accurately regions of the cardiovascular system \citep{chen2009effect, torii2009fluid, xiong2011simulation}. More recently an FSI study of a patient-specific fistula found that the effect of the wall compliance on the haemodynamics is non-negligible \citep{decorato2014numerical}. Their results showed that the velocities and WSS were overestimated by CFD simulations. 
	
	Phase-contrast magnetic resonance angiography (PC-MRA) has also been used in cardiovascular time-dependent blood flow visualisation and quantification \citep{markl2003time, grotenhuis2009validation}. Velocity encoding MRI techniques have shown promise in their ability to determine the velocity distribution of the flow and hence approximate parameters such as WSS and pressure changes and to locate regions of stenosis \citep{harloff20093d, frydrychowicz2009three}. MRI has also been used to produce valuable data for the boundary conditions required in numerical models and to compare the accuracy of the modelled and measured data \citep{gharahi2016computational, niemann2011computational, canstein20083d}.
	
	The objective of this research is the numerical investigation of the fluid dynamics and structural deformation of a patient-specific arteriovenous fistula. We model the blood flow through the anastomotic region of a fistula along with the deformation of the vessel walls by the partitioned coupling of two numerical methods, the finite element method (FEM) for the solid mechanics and the finite volume method (FVM) for the fluid dynamics. The numerical model is implemented in ANSYS\textsuperscript{\textregistered} Academic Research, Release 17.2, a widely-used commercial multiphysics simulation package. Specifically, we use ANSYS\textsuperscript{\textregistered} Fluent\textsuperscript{\textregistered} to model the blood flow and ANSYS\textsuperscript{\textregistered} Mechanical\texttrademark\ to model the vessel walls. The artery and vein are modelled with a hyperelastic material law, while it is assumed that the blood behaves as a Newtonian fluid. The modelling schemes and set relaxation parameters that optimise the stability and efficiency of the FSI model are assessed. The validated model is made available at \citep{guess2017model} for analysis and to be used as a basis for further investigation. The FSI and CFD models are quantitatively compared to determine the relative accuracy and efficiency between the two approaches. The ability of the PC-MRA technique to complement numerical modelling of vascular access is also assessed.

	\section{Partitioned fluid-structure interaction coupling}\label{StagFSI}
	Two dominant methods of FSI coupling exist. In the monolithic approach the coupling forms a single set of equations so that the fluid and solid models are solved simultaneously. Alternatively, in a partitioned (staggered) approach, the fluid and solid are solved as two separate systems sequentially. For strongly-coupled FSI problems a semi-implicit coupling scheme and relaxation techniques are required to mitigate the numerical instability inherent in partitioned FSI schemes known as the \textit{artificial added-mass effect}.
	
	The partitioned approach solves the system in coupling steps where each model is solved once and the solutions on the boundary transferred between the subsystems. For the problem at hand, where the forces on the solid are entirely dependent on the fluid solution, the fluid model should be solved first each coupling iteration. The General Grid Interface (GGI) algorithm \citep{Galpin1995} is used in transferring the fluid model boundary solution (the pressure) as the conservative variable force, from the fluid boundary mesh to the solid boundary nodes. To transfer the non-conservative solid boundary solution (the displacement) the Smart Bucket (SB) algorithm \citep{jansen1992fast} is used.

         \begin{figure}
             \begin{center}
                 \includegraphics[width=.8\textwidth]{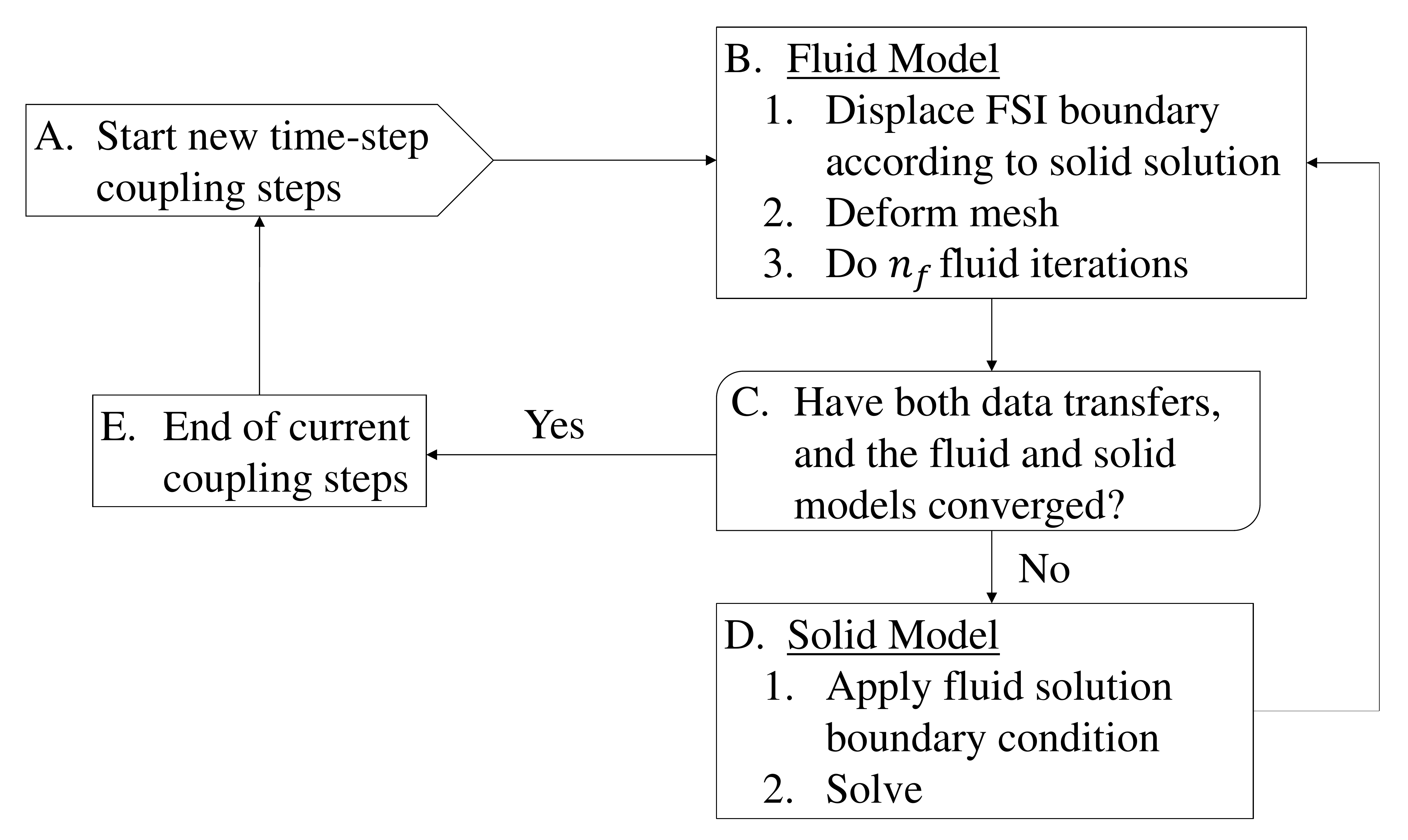}
             \end{center}
             \caption{Flow chart for the semi-implicit staggered FSI coupling algorithm}
             \label{fig:FSI}
         \end{figure}
         	
	Once the fluid model receives a boundary displacement solution update from the solid model a mesh smoothing operation is performed to deform the mesh so that its quality is maintained. Convergence of the FSI model is achieved when each of the solvers are converged and the change in each of the boundary solution transfers between coupling steps is considered negligible.
		\subsection{The added-mass effect and boundary stabilisation}\label{AddMass}
		Partitioned FSI coupling schemes suffer from an inherent numerical instability known as the added-mass effect, particularly in strongly coupled problems. To overcome this instability relaxation techniques are required to slow the convergence rate of the FSI model increasing its computational expense considerably. In addition to this, the increased computational expense in modelling the vessel walls each coupling step, make FSI models of strongly coupled cardiovascular systems take many times longer to simulate than the corresponding fluid model alone \citep{decorato2014numerical, ngoepe2011numerical, Degroote2008}.

		The added-mass effect occurs as a result of the solution, and boundary condition, mismatch between the partitioned models. This effectively results in the solid model displacing and entraining fluid so that the structural model appears to have `added-mass'. This `added-mass' acts on the structural degrees of freedom at the interface between the solid and fluid \citep{forster2007artificial}. A major issue with this numerical instability is that it increases as the time-step size decreases.

		A semi-implicit coupling scheme ensures that a number of easily implementable relaxation techniques can be used to overcome the added-mass instability. Using a small set number of fluid iterations each coupling step stabilises the FSI model by slowing down its convergence rate by solving the fluid model only partially between data transfers. The coupling procedure is illustrated in Figure~\ref{fig:FSI}.		
						
		Another effective stabilising technique in very strongly-coupled FSI problems is to modify the FVM continuity equation so that the diagonal entries of the linear matrix system are rescaled according to
		\begin{align}
		\vec{a} \leftarrow \vec{a} + KV\vec{I}
		\end{align}
		where $ K $ is the scaling factor and $ V $ is the volume of the cell adjacent to the boundary \citep{ANSYSFluentUsers}. Increasing the scaling factor $ K $ improves the diagonal dominance of the cells adjacent to the coupling interface. This method of stabilisation slows the convergence rate of the fluid primarily on the boundary such that the force transfer changes are more smooth without affecting the computational expense as much as other forms of relaxation.
		
\section{Patient-specific fluid model}\label{sec:Fluid}
	\subsection{PC-MRA acquisition and post-processing}\label{sec:MRA}
	The case studied is a patient's brachio-cephalic fistula. All data were acquired from a number of PC-MRA scans which took place at Groote Schuur Hospital (Cape Town, South Africa). These were performed with a 1.5T MRI scanner (MAGNETON Symphony Siemens AG, Erlangen). Images were sequenced and processed as described in \citep{Jermy}, and based on the investigational prototype 3D MRI velocity mapping acquisition methods presented in \citep{markl20124d}. The study was approved by the University of Cape Town Engineering and Built Environment Ethics in Research Committee.
	
	Initial scans were carried out on access patients recruited from the chronic haemodialysis program at Groote Schuur Hospital Renal Uni to first test and configure the PC-MRA and post processing technique. Patient scans were performed a minimum of 6 hours after dialysis treatment took place. No attempt was made however to control for relative hyperaemia, where higher blood flow rates than normal may arise after exercise or fluid consumption. 
	
	The 3D velocity encoded acquisitions were performed in a sagittal plane. The region of interest was positioned such that the patients' entire vascular access could be captured in one acquisition lasting between 20 and 30 minutes. Details of the sequences are shown in Table~\ref{tbl:MRIDet}.
	\begin{table}
		\centering
		\caption{PC-MRA sequence details}\label{tbl:MRIDet}
		\begin{tabular}{@{}cccc@{}}
			\toprule
			\multicolumn{1}{c}{Temporal resolution (\si{\sec})} & \multicolumn{1}{c}{$ v_{enc}$ (\si{\centi\metre\per\second})} & \multicolumn{1}{c}{Pixel spacing (\si{\milli\metre})} & \multicolumn{1}{c}{Slice thickness (\si{\milli\metre})}  \\ \midrule						
			0.0157       & 80        & 1.56 / 1.56               & 2                       \\\bottomrule
		\end{tabular}
	\end{table}
	
	\subsection{Geometry reconstruction and discretisation}\label{sec:MRAGeom}
\begin{figure} 
			\centering
			\begin{subfigure}{\textwidth}
			\includegraphics[width = \textwidth]{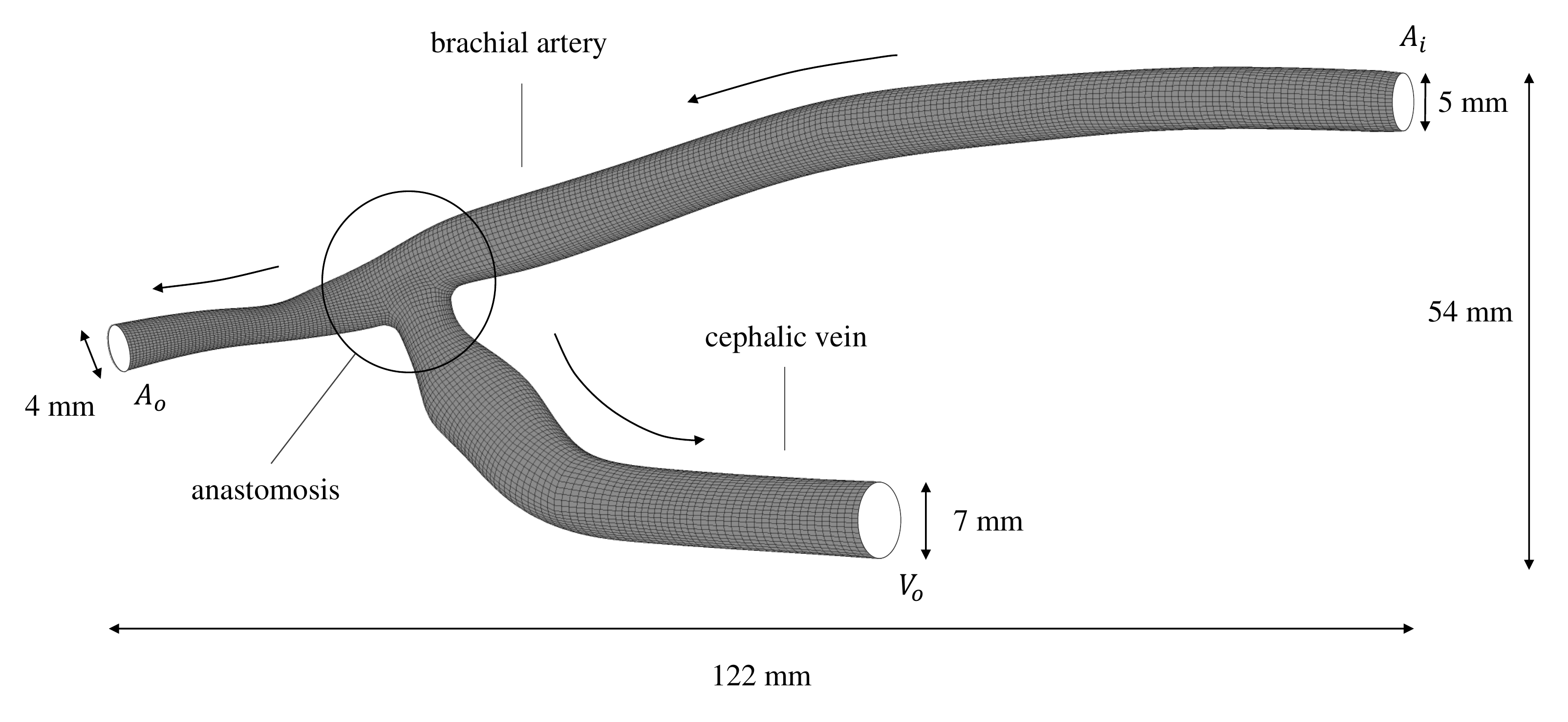}
			\caption{} 
			\label{fig:Fluid_dimensions}
			\end{subfigure}
			
			\begin{subfigure}{\textwidth}
			\centering
			\includegraphics[width = \textwidth]{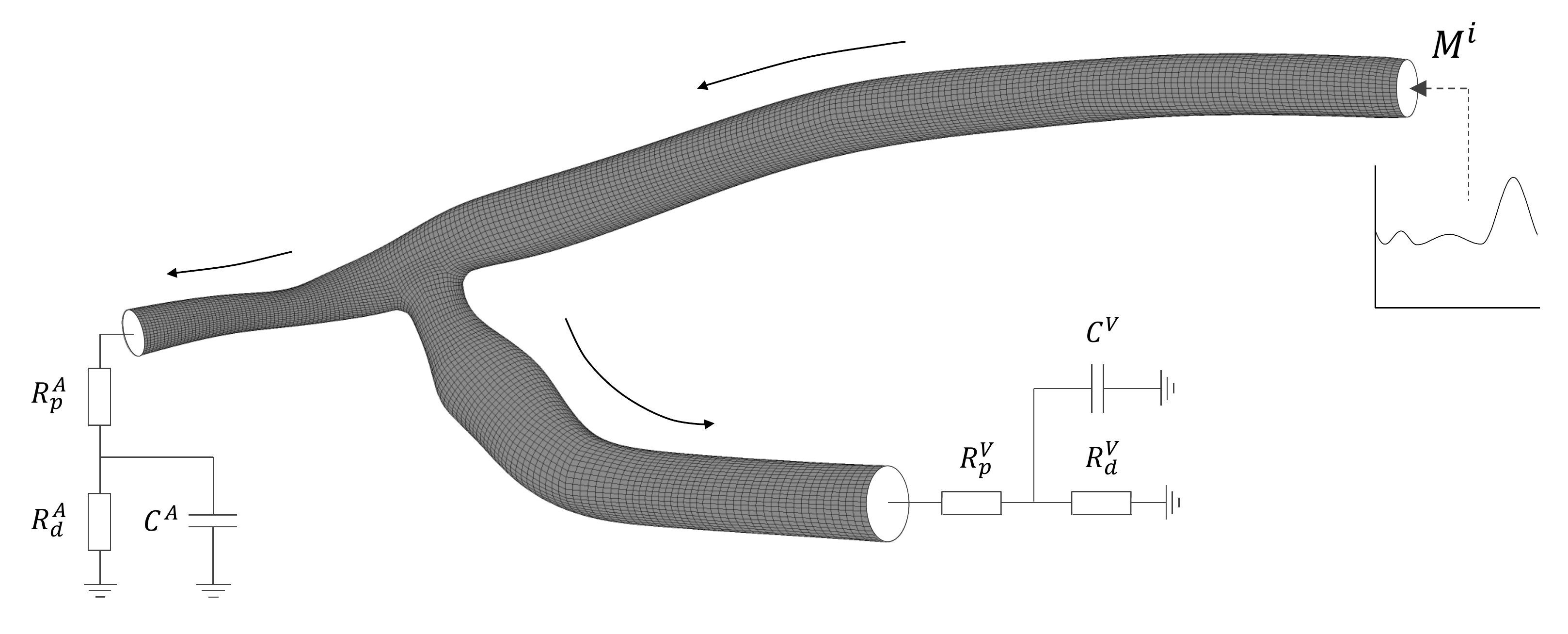}
			\caption{} 
			\label{fig:Fluid_bcs}
			\end{subfigure}
			\caption[Patient-specific fluid domain and boundary conditions]{Fluid domain and surface mesh of the patient-specific fistula: \subref{fig:Fluid_dimensions} details and dimensions, \subref{fig:Fluid_bcs} fluid model boundary conditions.} \label{fig:Fluid_Domain}
		\end{figure}

	Construction of the geometry required a substantial amount of smoothing since the spatial resolution of the 3D sequence was of the same order of the dimensions of the artery and vein. Meshlab, Solid Works\textsuperscript{\textregistered}, and ANSYS\textsuperscript{\textregistered} DesignModeler\texttrademark\  were used to construct and interpret the fluid domain for meshing. The point-cloud data was interpreted in Meshlab where the surface point-cloud was extracted. The smooth surfaces describing the limits of the fluid doamain and inner walls of the artery and vein were created from the extracted point-cloud surface in Solid Works. ANSYS DesignModeler software was used to interpret the geometry for use across ANSYS products and to extend the artery inlet length (see Section~\ref{sec:FluBCs}). The resultant domain and its proportions are shown in Figure~\ref{fig:Fluid_dimensions}. 

	The flow domain was discretised into an unstructured hexahedral mesh, visible in Figure~\ref{fig:Fluid_dimensions}, to minimise numerical diffusion and computational expense. This was achieved by a blocking approach that maximises the quality of the mesh and aligns the cells with the flow field as far as possible. This method also limits the development of skewed cells forming from fluid mesh deformation.
	
	\subsection{Boundary conditions and material behaviour}\label{sec:FluBCs}
	Blood flows in a pulsatile manner from the brachial artery upstream toward the anastomotic junction of the brachio-cephalic fistula. The blood is then primarily shunted through the cephalic vein, the low resistance pathway, back toward the heart. A smaller proportion of blood continues along the artery, the high resistance pathway, and flows along its natural path toward the downstream vessel structure. This behaviour and the upstream and down-stream influence on the blood flow through the fistula is replicated with an inlet mass flow rate waveform boundary condition and lumped parameter model outlet pressure boundary conditions.
	
\begin{figure}
		\centering
		\includegraphics[width = .65\textwidth]{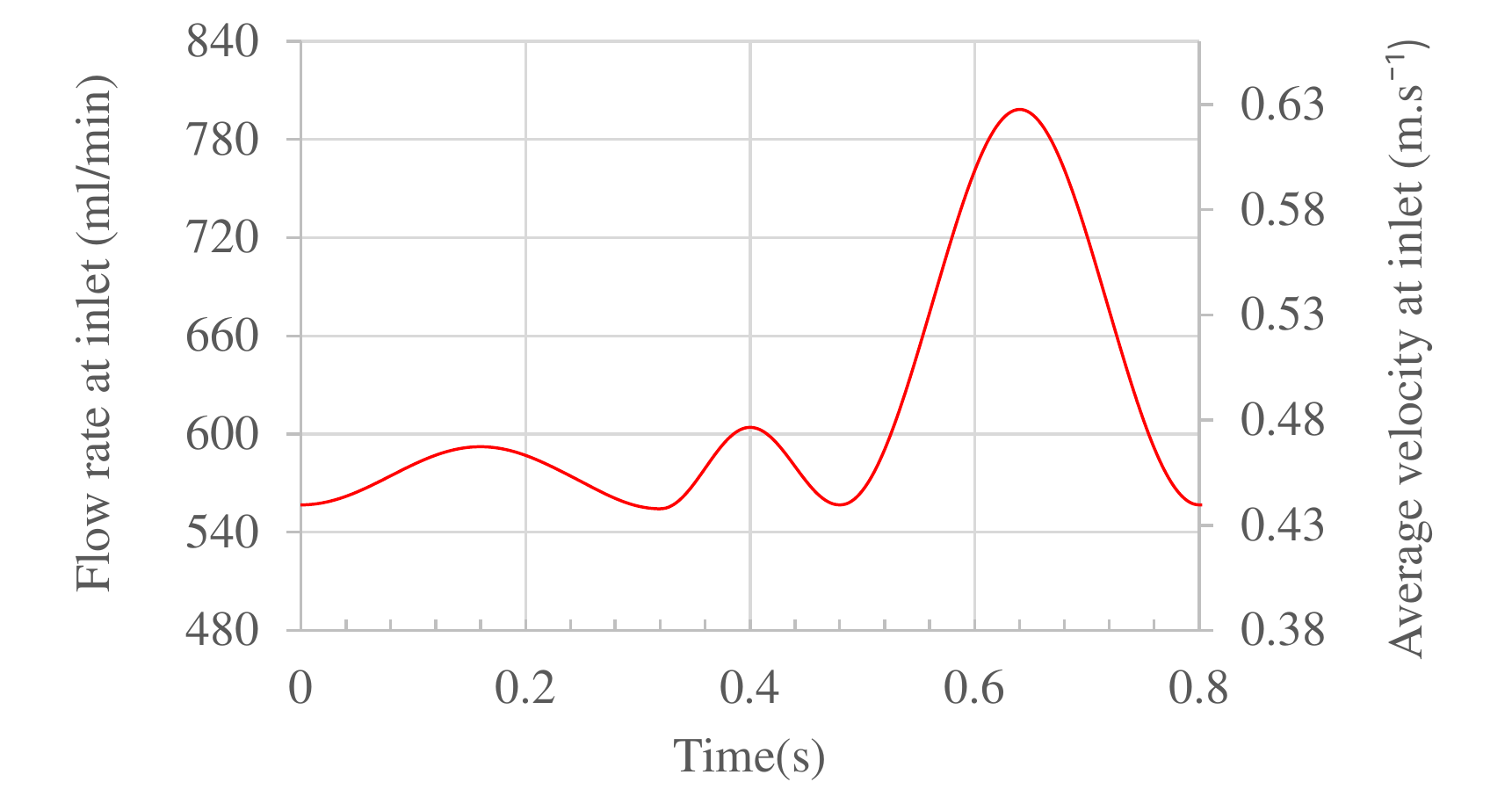}
		\caption[Artery inlet mass flow boundary condition]{The mass flow boundary condition waveform $ M_i $ prescribed at the artery inlet of the fluid model} 
		\label{fig:img_Inl_prof}
	\end{figure}

	The blood flow waveform, shown in Figure~\ref{fig:img_Inl_prof}, is applied as a mass flow rate inlet boundary condition $ M_i $ at the artery inlet of the model. This waveform has a cardiac-cycle period of \SI{0.8}{\second} with a time-averaged flow rate and average velocity of \SI{616}{\milli\litre\per\min} and \SI{0.49}{\metre\per\second} respectively. The pressure outlet boundary conditions are imposed by the three-element Windkessel lumped parameter model described in more detail in the following section. These boundary conditions are illustrated in Figure~\ref{fig:Fluid_bcs}. A no-slip condition is imposed on the walls of the vessel.
	
	Blood ordinarily exhibits Newtonian behaviour in the arterial system of the arm, where the viscosity remains relatively constant, as a result of high shear rates (of above \SI{100}{\per\second}). However, shear rates lower than \SI{100}{\per\second} are usually prevalent in the mature vein of an AV shunt where velocities are somewhat lower due to the bifurcation of flow and considerable enlargement of the vessel. This results in shear-thinning due principally to Rouleaux disaggregation in blood plasma \citep{essabbah1981transient}. Studies have found that the WSS generated in the enlarged veins of fistulae are under-predicted by a Newtonian model \citep{decorato2014numerical, shibeshi2005rheology}. The significant non-Newtonian behaviour is more accurately simulated by a Casson or Carreau-Yusada material model \citep{boyd2007analysis}. To limit the complexity of the fluid model we assume the blood is an isotropic, homogeneous, and incompressible Newtonian fluid with a density of \SI{1050}{\kilogram\per\metre\cubed} and a viscosity of \SI{3.2}{\Pascal\second}. 
		
	The assumptions and boundary conditions chosen result in a Womersley number and effective Reynolds number of 6.5 and 1124, respectively, in the artery upstream of the anastomosis. Laminar flow is thus predicted in the entire domain with a boundary layer thickness of approximately \SI{0.6}{\milli\metre}. Under these assumptions the flow profile is predicted to develop fully \SI{98}{\milli\metre} from the inlet \citep{hale1955velocity,he1994unsteady}. The entrance length was thus of sufficient length (greater than \SI{100}{\milli\metre}) to ensure that a fully-developed flow profile is established before reaching the anastomosis.

		\subsubsection{The three-element Windkessel lumped parameter model}\label{sec:Wind}
		Lumped parameter models have been used extensively in modelling the cardiovascular system as electric circuit analogues, particularly the system of vessels downstream of the domain of interest which would otherwise be impractical \citep{kim2009coupling,kim2010patient}. An RCR three-element Windkessel model implemented at the outlets to mimic influence of the upstream vessel resistance, and the downstream vessel resistance and compliance, on the flow through the fistula. 

		The electric circuit analogue of the RCR Windkessel model is a resistor in series with a resistor and a capacitor connected in parallel. The lone resistor $ R_p $ acts as the proximal resistance (upstream resistance). The resistor and capacitor in parallel act as the distal resistance and compliance of the downstream vascular bed. The differential equation of the three-element RCR Windkessel model is given by
		\begin{align}
		\dfrac{\partial p_o}{\partial t} 	 = R_p \, \dfrac{\partial Q_o}{\partial t} + \dfrac{p_o}{\tau} + \dfrac{p_o\,(R_p+R_d)}{\tau},
		\label{eq:WindDif}
		\end{align}			
		where $ R_p $ and $ R_d $ are the proximal and distal resistance, $ Q_o $ is the volumetric flow rate at the outlet, and $ p_o $ is the average static pressure at the outlet \citep{westerhof2009arterial}. The time constant $ \tau $ describes how quickly the system responds to a changes in flow rate and is given by $ \tau = R_d \, C $. The state-variable form of the Windkessel equation is given by
		\begin{align}
		p_o(t) = \big[p_0 - R_p\, Q_o(0)\big]\,e^{-t/\tau} + R_p\, Q_o + \int_{0}^{t}\dfrac{\exp[(\tilde{t}-t)/\tau]}{C}Q_o(\tilde{t})\,\d \tilde{t}.
		\label{eq:WindInt}
		\end{align}
		The state variable form takes advantage of the viscoelastic like behaviour of the Windkessel model so that the difficulty in retaining history data in the numerical scheme is avoided. Dropping the first term in the equation and discretising in time implicitly by the Crank-Nicolson method, we obtain
		\begin{align}
		p_o^{t+\Delta t} = R_p\,Q_o^{t+\Delta t} + \dfrac{h^{t+\Delta t}}{C},
		\label{eq:WindNum}
		\end{align}
		where $ h^{t+\Delta t} $ is given by 
		\begin{align}
		h^{t+\Delta t} = \frac{\Delta t}{2}\,Q_o^{t+\Delta t} + \exp\bigg(\dfrac{\Delta t}{\tau}\bigg)\left( h^{t}+\dfrac{\Delta t}{2}\,Q_o^{t}\right).
		\label{eq:WindHnp1}
		\end{align}
		
		The simulation needs to carry out two cardiac-cycles before becoming sufficiently periodic (error inferior to \SI{1}{\percent}). The term $ \big[p_0 - R_p\, Q_o(0)\big]\,e^{-t/\tau} $ in equation (\ref{eq:WindInt}) can therefore be safely neglected since it becomes negligible in less than one and a half cardiac-cycles. Lumped parameter type boundary conditions are not a built-in feature of  ANSYS Fluent software and are implemented here using a ``User-Defined Function'' that is executed by ANSYS Fluent software each iteration.
		
		The pressure and flow rate waveforms at the inlet, or at one of the outlets, as well as the approximate flow-rate split throughout the model are required to adequately calibrate the Windkessel models. Windkessel models, and more complex lumped parameter models, can be calibrated to a higher accuracy with pressure and flow rate waveforms at each of the inlets and outlets.
		
		Blood pressure measurements are typically taken with a sphygmomanometer, this determines the systolic and diastolic blood pressures in the upper arm. These pressure readings cannot be performed on a dialysis patient's on the arm with an access site since restricting blood flow may lead to access trauma \citep{kdoqi2006}. Invasive measurements such as this are highly inaccurate for determining day-to-day loading conditions. 
		
		To our knowledge, pressure data for a brachio-cephalic fistula is currently available. Measuring techniques with an intravascular catheter have, however, been used post operatively to determine the blood pressure distal to a Brachio-Cimino fistula \citep{corpataux2002low}. This method is also invasive but could potentially provide adequate readings of pressure proximal or distal to the anastomosis of a fistula or graft. The viability of this technique, or a non-invasive technique, to accurately pressure data for boundary condition to vascular access requires testing. 
		
		We follow the convention used by Decorato et al. \citep{decorato2014numerical}, whereby the Windkessel model at each of the outlets is calibrated to induce a flow split between the arterial and venous outlets of \SI{30}{\percent} - \SI{70}{\percent}, and to generate inlet pressures that range between 50 and \SI{65}{\mmMercury} (approximately \num{6.5} and \SI{8.5}{\kilo\Pascal}). The proximal and distal resistance, capacitance, and time-constant for each outlet RCR Windkessel boundary condition of the model are given in Table~\ref{WindValues}. 
		
		\begin{table}
			\centering
			\caption{RCR Windkessel outlet boundary condition model details}\label{WindValues}
			\begin{tabular}{lcccc}
				\toprule
				Outlet & \multicolumn{1}{c}{ ${R_p}$ (\si{\Pascal\second\per\metre\cubed})}    & \multicolumn{1}{c}{ ${R_d}$ ( \si{\Pascal\second\per\metre\cubed})} & \multicolumn{1}{c}{C(\si{\metre\cubed\per\Pascal})}& \boldmath$\tau$(\si{\second})  \\ \midrule	
				Artery & \num{1.12e9} & \num{1.12e9} & \num{1.34e-10} & 0.15 \\
				Vein  & \num{4.6e8} & \num{4.6e8} & \num{1.74e-10} & 0.08 \\ \midrule						
			\end{tabular}
		\end{table}			
	
	\subsection{Mesh refinement analysis and solver settings}
\begin{figure} 
		\begin{subfigure}{.5\textwidth}
			\includegraphics[width=\linewidth]{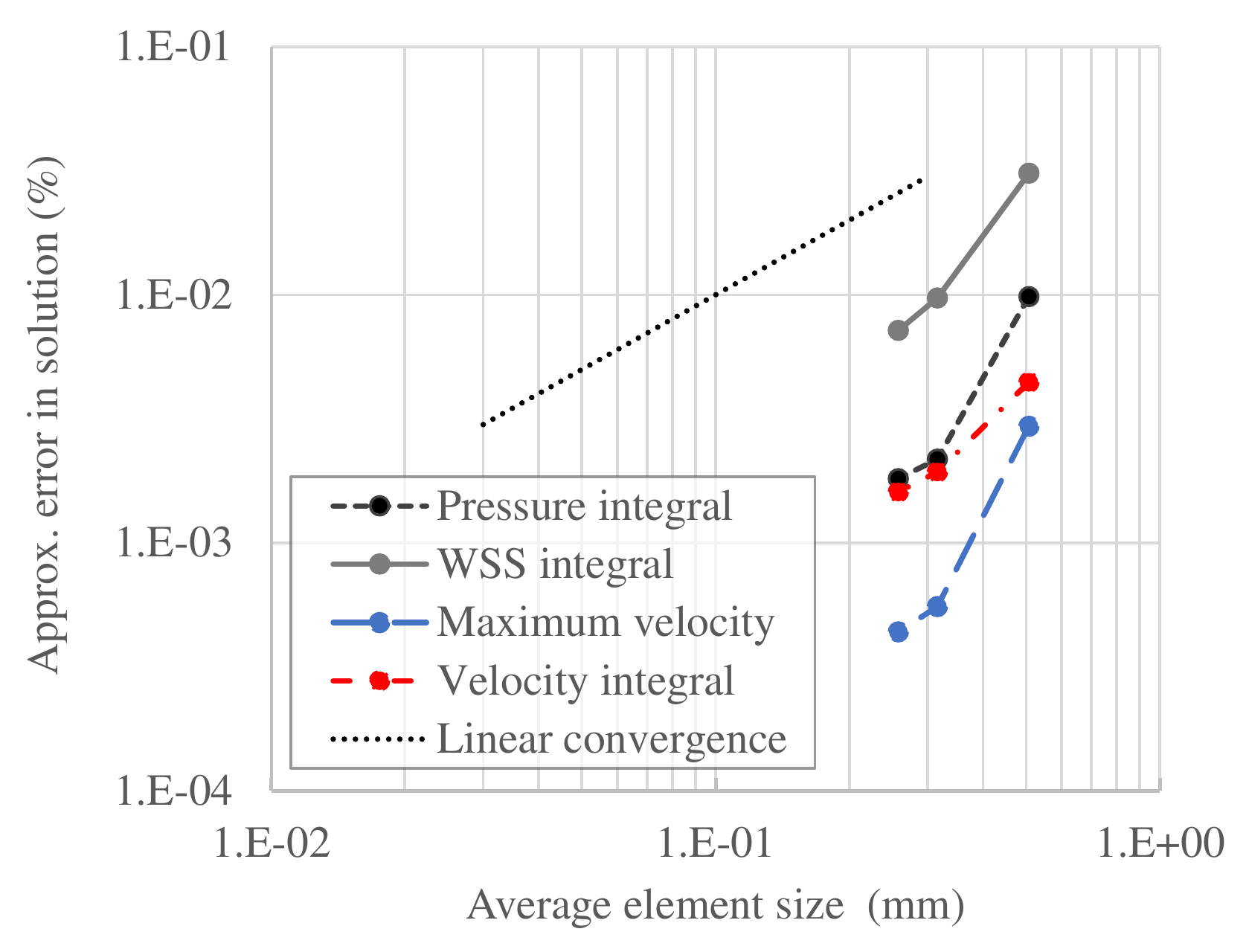}
			\caption{}\label{fig:FluidConv_loglog} 
		\end{subfigure}
		\begin{subfigure}{.5\textwidth}
			\includegraphics[width=\linewidth]{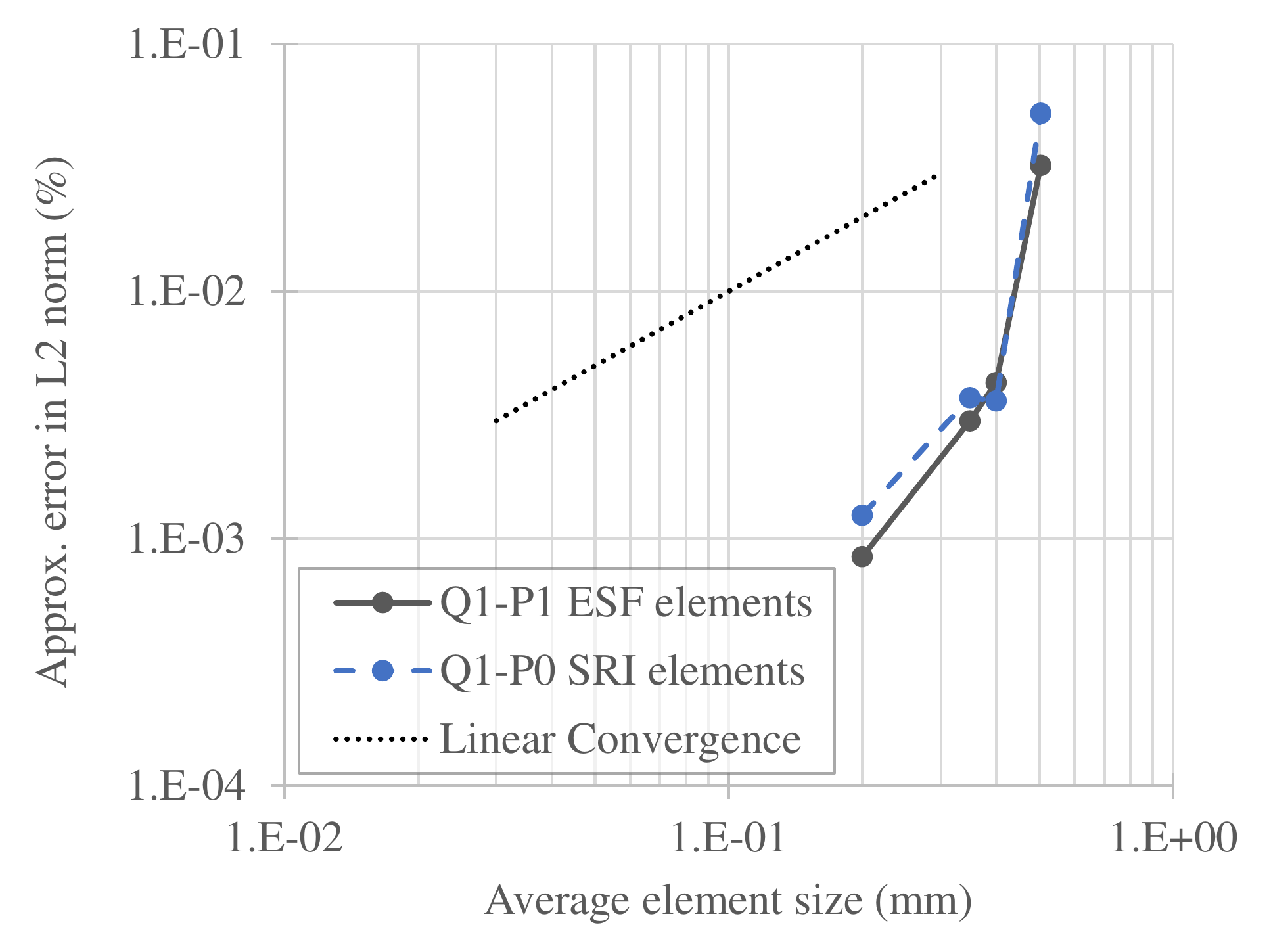}
			\caption{}\label{fig:SolidConv_loglog} 
		\end{subfigure}\hspace*{\fill}
		\caption[Fluid and solid mesh refinement convergence plot]{Mesh refinement convergence plots: \subref{fig:FluidConv_loglog} fluid mesh solution convergence and \subref{fig:SolidConv_loglog} solid mesh solution convergence for two linear element formulations} \label{fig:ConvPlots}
	\end{figure}

	A mesh refinement analysis carried out to ensure that the solution is mesh-independent and to find an optimal mesh choice for efficiency while achieving the required accuracy. The latter is determined by finding at which point the mesh is “sufficiently fine”; that is, where further refinement leads to negligible improvement. Since the solid and fluid meshes are made conformal at the FSI boundary (see Section~\ref{sec:GeoMesh}), this also serves to minimise the size of the solid problem. Three meshes were tested against a very fine reference mesh, with over a million nodes, in terms of the integrals of pressure, WSS, and velocity as well as the maximum velocity within the flow domain. The solution is mesh-independent and achieves the required accuracy (error inferior to 1\%) with a mesh of almost 200,000 nodes (average element size of \SI{0.316}{\milli\metre}) as shown in Figure~\ref{fig:FluidConv_loglog}.

	\subsection{Solver settings and solution behaviour}\label{sec:FluSet}
	A time-step size of \SI{0.005}{\second} was chosen to achieve the required temporal accuracy and solution convergence while minimising the computational expense of the problem. The PISO pressure-velocity coupling scheme \citep{issa1986solution} was chosen for its efficiency which performed best for our model with Skewness Correction and Skewness-Neighbour Coupling deactivated. The under-relaxation factors (URFs) were reduced from their typical values and chosen carefully to address the increased instability caused by the Windkessel boundary conditions and to maximise solver efficiency and accuracy. At a sufficiently small time-step size of \SI{0.005}{\second} the solver required 44 PISO iterations on average to converge each time-step with the solver discretisation and parameter settings summarised in Table~\ref{tbl:flu_set_Conv}. A single cardiac cycle requires less than 11 minutes to complete on an Intel\textsuperscript\textregistered Xeon(R) E5 workstation with 16 cores (12 used) and 64 GB RAM.
	
	\begin{table}[]
		\centering
		\caption{Optimal fluid solver discretisation and parameter settings}
		\label{my-label}
		\begin{tabular}{lr}
			\toprule
			\multicolumn{2}{c}{PISO scheme settings}                   \\ 
			Skewness Correction                    & 0                          \\
			Neighbour Correction                    & 1                          \\ 
			Skewness-Neighbour coupling            & Off                        \\ \midrule
			\multicolumn{2}{c}{Discretisation}                \\ 
			Pressure                               & 2nd-order upwind           \\ 
			Momentum                               & 2nd-order                  \\ 
			Time                                   & 2nd-order implicit         \\ \midrule
			\multicolumn{2}{c}{Under-relaxation factors ($ \Delta $t = 0.005s)} \\ 
			Pressure                               & 0.20                        \\ 
			Momentum                               & 0.47                      \\ \bottomrule
		\end{tabular}\label{tbl:flu_set_Conv}
	\end{table}

\section{Patient-specific FSI model}\label{sec:fsi}			

\subsection{Blood flow model}\label{sec:FluidFSI}
	The fluid model requires two adjustments to function in a staggered FSI configuration. The boundary maintains a no-slip wall condition but must be set to receive nodal displacements from the solid boundary solution at the start of each coupling step. Mesh movement is required to maintain a high quality mesh, especially when large deformations occur and/or the mesh is fine.  

	The spring-based mesh movement method was chosen where the edges of each cell are temporarily modelled as springs. This is the most efficient method for mesh movement available and maintains the high quality of the mesh throughout the simulation cycle. The ``Laplace node relaxation'' parameter is set to unity, or near unity, to ensure that the internal nodes move correctly in relation to the boundary nodes.
\subsection{Vessel wall model}\label{sec:Solid}
	\subsubsection{Geometry and mesh setup}\label{sec:GeoMesh}
	The resolution of the MRI sequences would need to be greatly increased to segment the vessel walls. As such, the thickness of the vein and artery can not be determined from the MRI data. It is assumed that the vein has a constant thickness of \SI{0.4}{\milli\metre}, a thickness typical for a cephalic vein prior to AVF creation \citep{corpataux2002low}. The arterial thickness is set to 1/10th of the lumen diameter, the approximate ratio according to \citep{gutierrez2002automatic}. The geometry and mesh of the vessel walls were generated by extruding the fluid boundary mesh. This was done, as opposed to using a shell element mesh, to incorporate a Robin boundary condition on the outer walls of the artery and vein (see Section~\ref{sec:solidBCs}). The change in thickness between the artery and vein was created by tapering the mesh during extrusion in ANSYS\textsuperscript{\textregistered}ICEM CFD\textsuperscript{\texttrademark} software.

	\subsubsection{Material model}\label{sec:MatModel}
	The vessel walls are made up of a complex set of layers of tissue with helically distributed collagen fibres \citep{gasser2006hyperelastic}. Increased tensile stress and deformation cause these fibres to reorientate in the direction of the principal strain so that the tensile strength of the vessel increases primarily in the direction of principal strain. Anisotropic hyperelastic material models are able to model this tissue and fibre behaviour. However, one can only apply this material model to perfectly cylindrical domains in ANSYS Mechanical software. It is currently not possible to apply fibre orientations to a complex domain such as a patient specific fistula, though it is possible to do this loosely when combined with ANSYS\textsuperscript\textregistered\ Composite PrepPost software. The viability or accuracy of this method was not tested.
	
	We have assumed that the vessel walls follow the third-order Yeoh model \citep{yeoh1993some} where the mechanical properties are differentiated between the artery and vein. Material constants were obtained from \citep{decorato2014numerical} which have been determined form the experimental data of \citep{mcgilvray2010biomechanical} for the venous tissue, and of \citep{prendergast2003analysis,maher2011inelasticity} for the arterial tissue, the latter having a larger compliance.
	
	\subsubsection{Boundary and initial conditions}\label{sec:solidBCs}
	The model is constrained each of the ends of the domain by Dirichlet boundary conditions. The inlet is fixed in place fully to aid in stabilising the final FSI model where the pressure, and pressure changes, are greatest. This unphysical condition has negligible effect on the overall solution since and is far from the anastomosis and vein the region of particular interest. At the outlets, nodes are prevented from displacing axially but are free to translate parallel with the surface made by each outlet. These frictionless type boundary conditions allow the outlets to deform naturally due to the internal pressure.
	
	\begin{figure}
		\centering
		\includegraphics[width = \textwidth]{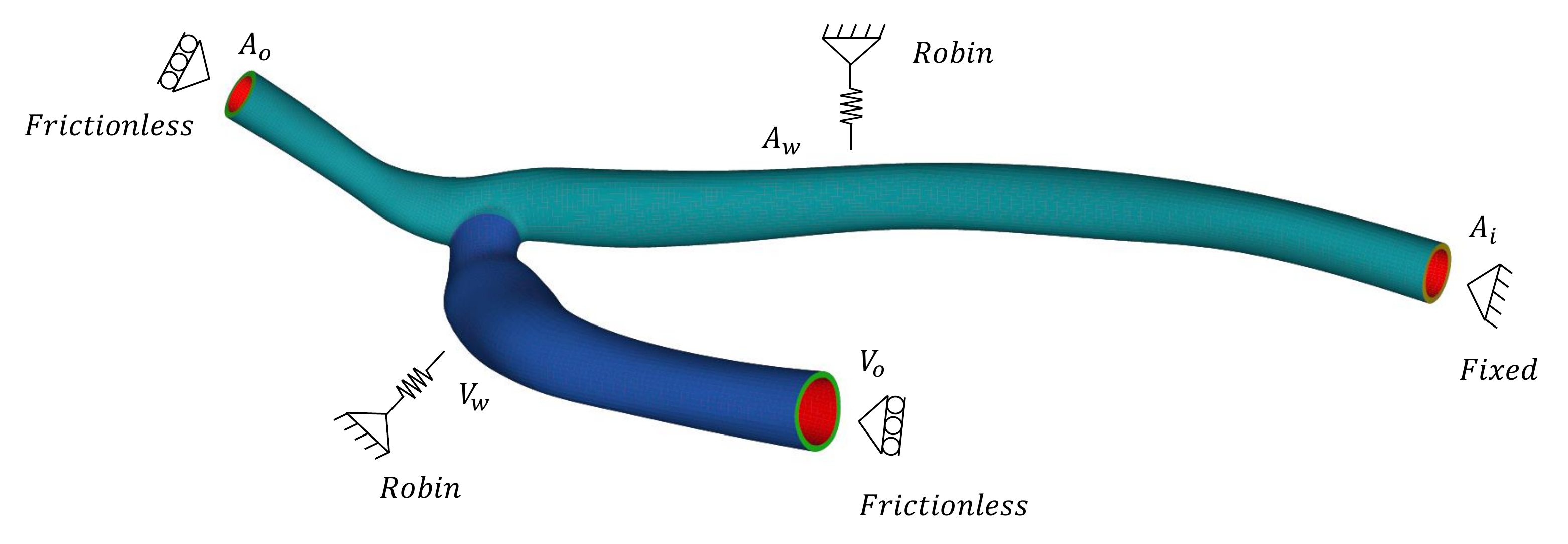}
		\caption[Solid model boundary conditions]{Vessel wall model boundary conditions}
		\label{fig:img_SolidBCs}
	\end{figure}
		
	A Robin boundary condition has been implemented on the outer surfaces of the vessel walls to mimic the behaviour of the tissue surrounding the vessel walls. No data could be found on the typical stiffness characteristics of the tissue surrounding the veins and arteries in the upper arm. The stiffness of the Robin boundary condition was thus set at $ k= $ \SI{0.01}{\newton\per\milli\meter\cubed} to  constrain the model sufficiently but have negligible effect on the strain in the vessel walls. Cardiovascular vessels, such as those in vascular access, are at all times in vivo in a stressed state due to the continuous pulsatile pressure generated by the heart. The effects of prestressing have however not been considered in the present study. The set of boundary conditions applied to the solid model is depicted in Figure~\ref{fig:img_SolidBCs}. The artery inlet, outlet and outer walls are indicated by $ A_i $, $ A_o $, and $ A_w $ respectively, while the vein outlet and outer walls are indicated by $ V_o $ and $ V_w $.

	\subsubsection{Mesh sizing and element choice}\label{sec:ELem}	
	The performance of a number of element types and formulations in modelling the vessel walls were examined with respect mesh size. Four base mesh sizes were studied with each element type to ensure that the solid solution was mesh independent and to determine the best combination of element choice and mesh size for solution efficiency and accuracy. Elements designed to circumvent locking in the incompressible limit were considered and include the following:
	\begin{itemize}
		\item Q1-P0 hexahedron with 1 pressure degrees of freedom (DOF) and selective reduced integration (SRI) where shear strains or volumetric strains are evaluated in the element midpoint only.
		\item Q1-P1 hexahedron with the enhanced strain formulation (ESF) with 4 pressure DOF and a further 9 enhanced strain DOF.
		\item Q$^\prime$2-P1 serendipity 20 noded hexahedron with a further 2 internal nodes and 4 pressure DOF.
	\end{itemize}
	These standard elements are provided by ANSYS Mechanical (for more details see the ANSYS Mechanical APDL Documentation \citep{ANSYSAPDLTheory}).
	
	The use of the Q$^\prime$2-P1 serendipity element produced the most accurate results and fastest converging solution with mesh refinement. The reference solution was thus chosen to be the one produced from the highest density Q$^\prime$2-P1 mesh containing over 3.5 million DOFs. The ESF elements predicted the solution more accurately (error inferior to 1\%) than the SRI elements when a mesh containing almost 100,000 DOFs was used. The solution was found to be mesh independent (see Figure~\ref{fig:SolidConv_loglog}) and the computational expense minimised using ESF elements with a single layer of elements through the thickness the vessel walls.

	\subsection{Numerical set-up and behaviour}\label{sec:NumSet}
	This primary form of stabilisation of the added-mass effect comes from using a very low number of iterations per coupling step so that the fluid solution only converges to a small degree between boundary solution transfers. The URFs can also be lowered to essentially perform the same form of stabilisation, by slowing the fluid convergence globally (across its domain). This method is less effective at FSI stabilisation than lowering the number of iterations per coupling step however.
	
	The boundary source coefficient stabilisation method described in Section~\ref{AddMass} slows the convergence rate of the fluid solution primarily on the FSI boundary. This provides an effective form of stabilisation that does not slow the overall FSI solution rate as much as the prior methods. Boundary stabilisation, however, interacted poorly with the Windkessel outlet models and destabilised the fluid model. We found that one could use this stabilisation effectively if the URFs are decreased as the degree of boundary stabilisation is increased. A combination of these stabilisation methods thus optimally stabilises the added-mass effect and maximises the efficiency of the overall model.

	Due to the low stability of the fluid iterations at the start of each time step, in part due to the Windkessel outlet models, it is unnecessary for the solid solution (which converges fully each coupling step) to closely follow the fluid solution initially. This was achieved by ramping the transferred pressure solution to the solid model over the first few coupling steps. This implementation also had a stabilising effect on the overall FSI model and reduced the number of coupling steps required to converge each time step.
	
		\begin{table}
			\centering
			\caption{Fluid and FSI model settings for optimal stabilisation and efficiency}\label{tbl:optFSI}
			\begin{tabular}{lccc}
				\toprule
				\multicolumn{1}{l}{Parameter} & \multicolumn{1}{c}{Fluid (dt = 5 ms)} & \multicolumn{1}{c}{FSI (dt = 5 ms)} & \multicolumn{1}{c}{FSI (dt = 2 ms)}                 \\ \midrule
				Pressure URF:                             	& 0.20  & 0.10    & 0.09   \\
				Momentum URF:                             	& 0.47  & 0.23     & 0.21  \\
				Boundary stabilisation scale factor   		& n/a  	& 200     & 180      \\
				No. of ramped coupling iterations       	& n/a   & 6    & 6        \\
				No. of coupling steps per time step      	& n/a	& 22   & 17       \\  
				No. of fluid iterations per coupling step   & n/a   & 5    & 5     \\ 
				No. of fluid iterations to converge fully   & 44  	& 115    & 85      \\ \bottomrule	             
			\end{tabular}
		\end{table}

	A combination of all the stabilising methods described above thus optimally stabilise the model. The parameters that accomplished this and maximised the efficiency of the patient-specific FSI model, with time step size of 0.005s, are shown in Table~\ref{tbl:optFSI}. This Table includes the optimal parameters for the corresponding uncoupled patient-specific fluid model. The computational expense maintains a linear indirectly proportional relationship to the time-step size over the range of time-steps tested (see Table~\ref{tbl:timeFSI}). The FSI models were run on a Xeon(R) E5 workstation with 16 cores and 64 GB RAM, the fluid model was run on twelve of the cores and the solid model on four.
	
		\begin{table}[]
			\centering
			\caption{Computational time per cardiac-cycle for the FSI model}\label{tbl:timeFSI}
			\begin{tabular}{@{}cccc@{}}
				\toprule
				\multicolumn{1}{c}{Time-step} & \multicolumn{1}{c}{No. of} & \multicolumn{1}{c}{Coupling iterations} & \multicolumn{1}{c}{Total CPU}  \\
				\multicolumn{1}{c}{size} & \multicolumn{1}{c}{time-steps} &\multicolumn{1}{c}{per time-step} & \multicolumn{1}{c}{time per pulse}  \\ \midrule						
				0.01 s        & 80        & 25             & 11 hrs                       \\ 
				0.005 s       & 160        & 23               & 20 hrs                       \\ 
				0.002 s       & 400        & 17               & 38 hrs                   \\ \bottomrule
			\end{tabular}
		\end{table}	

\section{Results and Analysis}\label{sec:FSIResults}				
\begin{figure}
		\centering
		\begin{subfigure}{.48\textwidth}
			\centering
			\includegraphics[height=.76\textwidth]{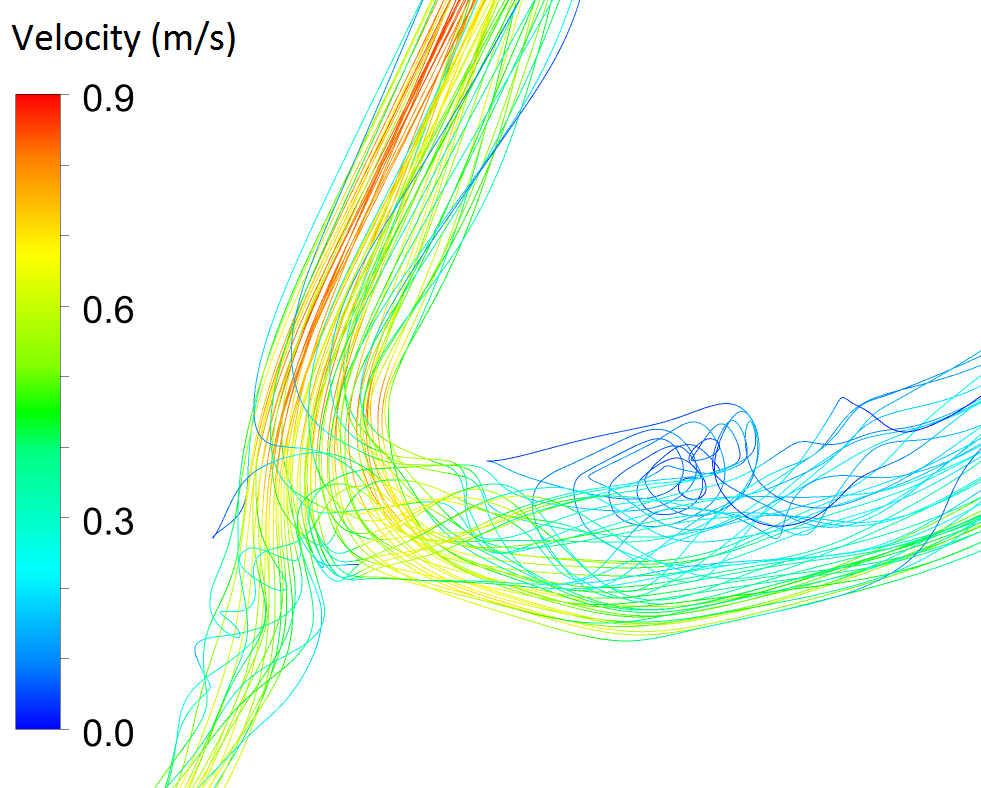}
			\caption{Fluid model} \label{fig:Fluid_Stream_Recirc}
		\end{subfigure}\hspace*{\fill}
		\centering
		\begin{subfigure}{.48\textwidth}
			\centering
			\includegraphics[height=.76\textwidth]{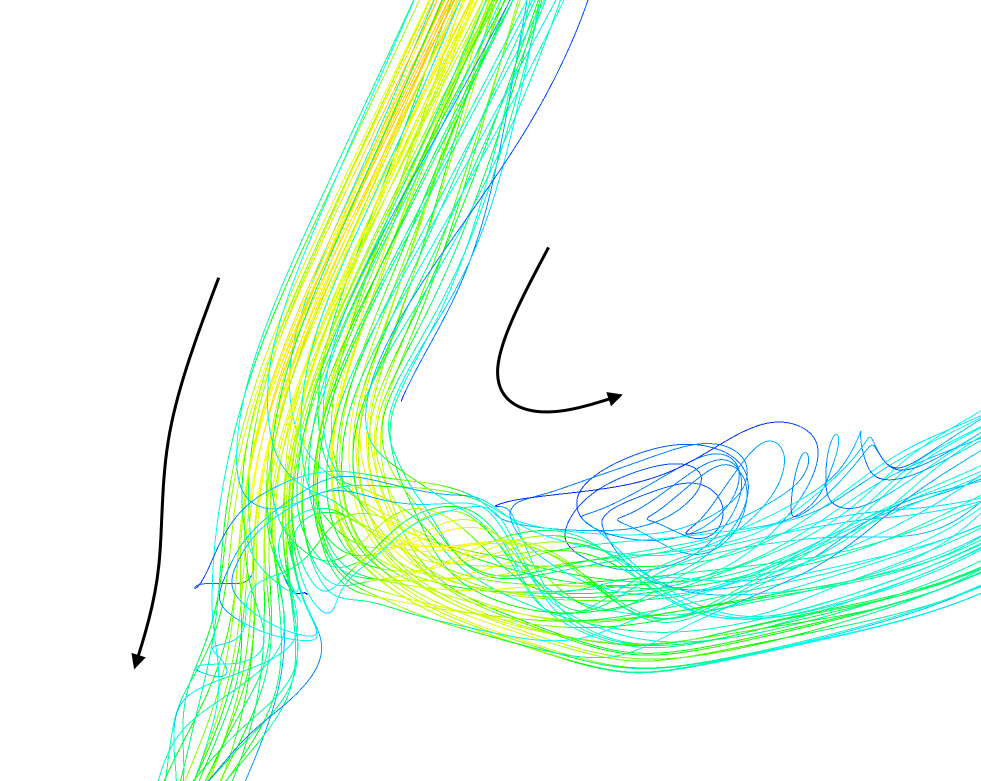}
			\caption{FSI model} \label{fig:FSI_Stream_Recirc}
		\end{subfigure}\hspace*{\fill}				
		\caption[Fluid and FSI model velocity streamlines]{Comparison of the velocity streamlines at the anastomosis and recirculation zone of the fluid and FSI models}\label{fig:vel_comp3}
	\end{figure}

\begin{figure} 
		\begin{subfigure}{.5\textwidth}
			\includegraphics[width=\linewidth]{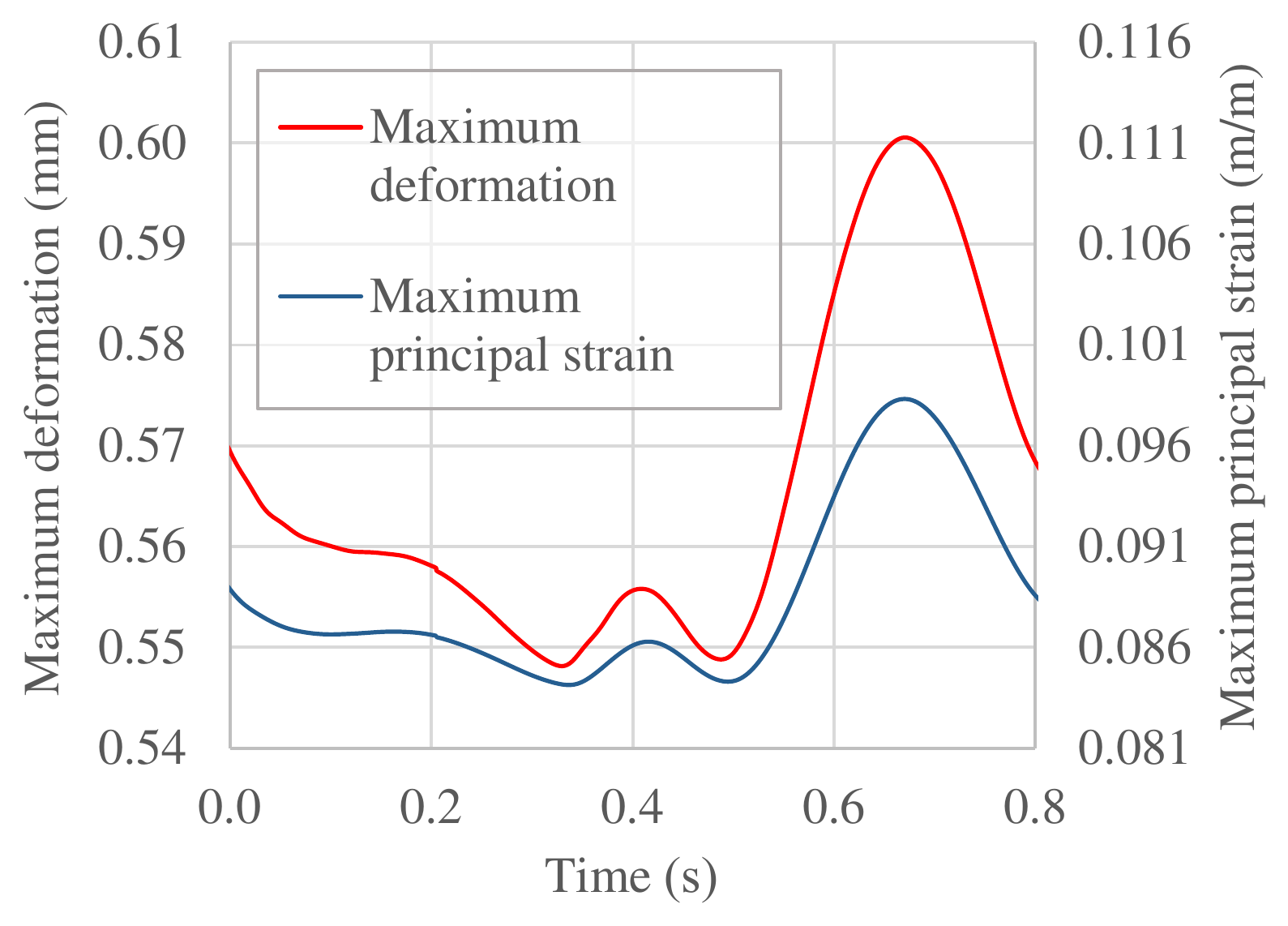}
			\caption{}\label{fig:MaxStrainDefArtery} 
		\end{subfigure}
		\begin{subfigure}{.5\textwidth}
			\includegraphics[width=\linewidth]{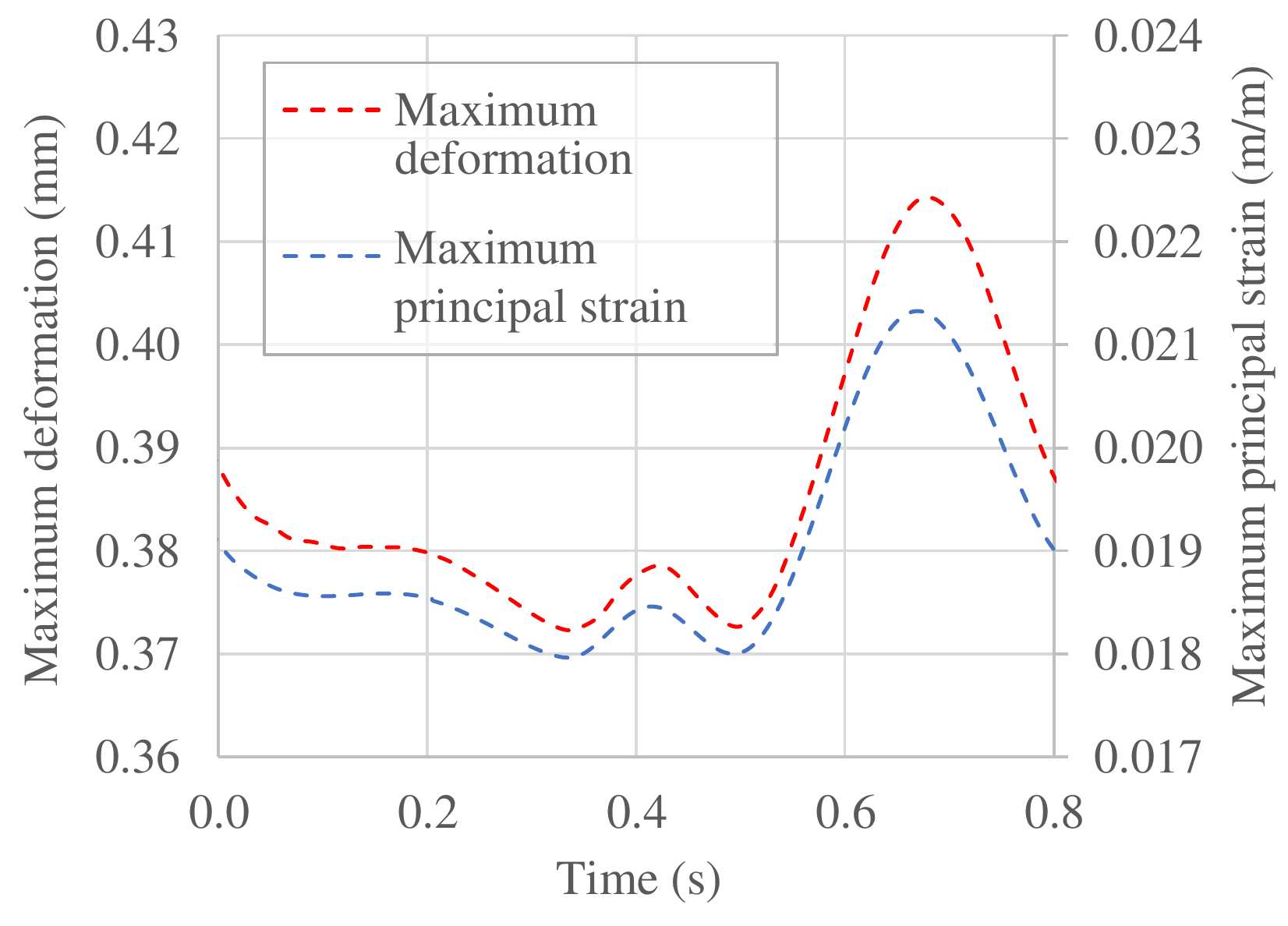}
			\caption{}\label{fig:MaxStrainDefVein} 
		\end{subfigure}\hspace*{\fill}
		\caption[Maximum deformation and principal strain profiles in the vessel walls]{Transient profiles of the maximum deformation and maximum principal strain predicted by the FSI model: \subref{fig:MaxStrainDefArtery} in the artery walls and \subref{fig:MaxStrainDefVein} in the vein walls} \label{fig:MaxStrainDef}
	\end{figure}

	The velocity streamlines of the fluid and FSI models are shown in Figure~\ref{fig:vel_comp3}. The fistula is viewed here to focus on the anastomosis, and the blood recirculation region in the vein at the bottom right of the figure. The artery here runs from the top of the figure to the bottom left with the blood flowing in the same direction. The flow features predicted by both models are almost identical. It is however evident that the maximum velocities in the artery are lower in the FSI model. This is because the expansion of the fistula increases the internal volume of the blood flow domain thereby reducing the flow velocity, particularly in the more compliant artery. However, the maximum strain and deformation are shown to be positive and non-negligible throughout the cardiac cycle in Figure~\ref{fig:MaxStrainDef}. This occurs because the solid model is not prestressed resulting in an overestimation of the strain, deformation, and velocity solutions throughout the cardiac-cycle. The FSI solution should return to the undeformed state (the initial prestressed state) each cardiac cycle.

\begin{figure}
		\begin{subfigure}{.5\textwidth}
			\centering
			\includegraphics[width=.9\textwidth]{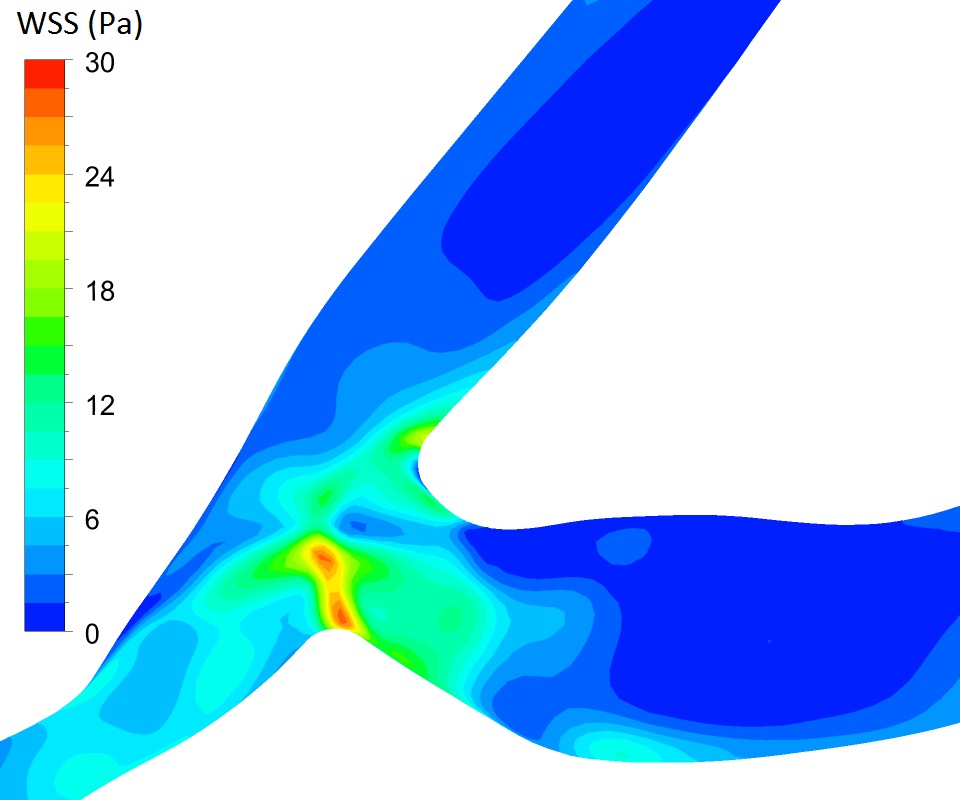}
			\caption{Fluid model} \label{fig:WSSCompFluid}
		\end{subfigure}
		\begin{subfigure}{.5\textwidth}
			\centering
			\includegraphics[width=.9\textwidth]{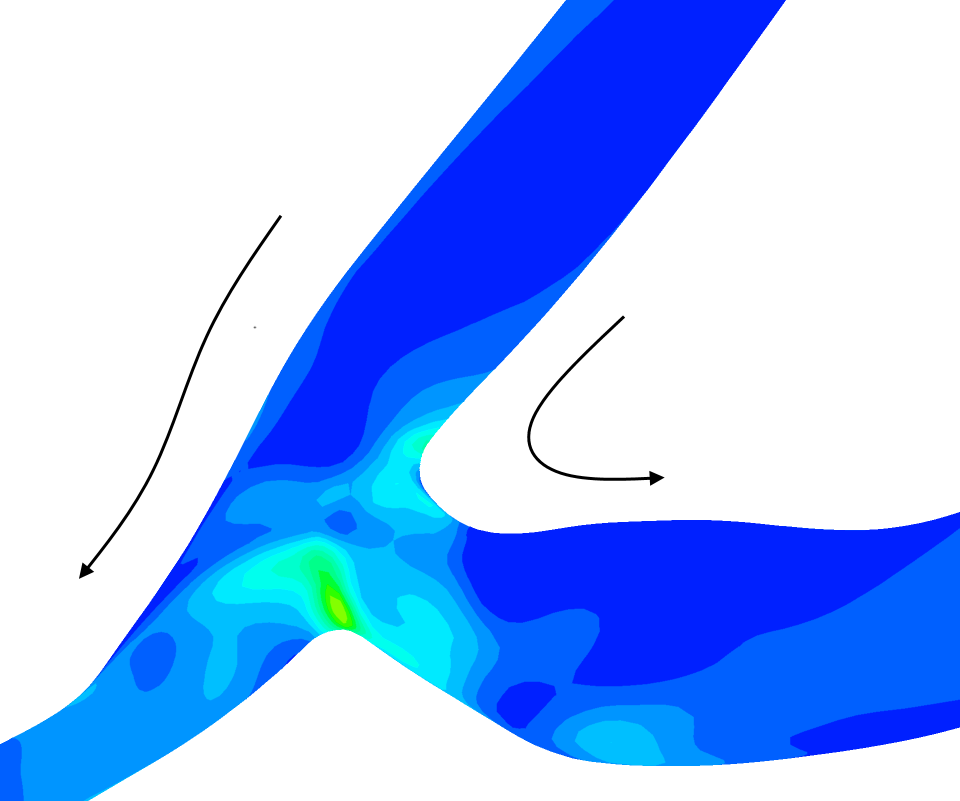}
			\caption{FSI model} \label{fig:WSSCompFSI}
		\end{subfigure}
		\caption[Fluid and FSI WSS solution contours]{Comparison of WSS contours between the fluid and FSI models}\label{fig:WSSComp}
	\end{figure}

\begin{figure} 
		\begin{subfigure}{.48\textwidth}
		\centering
			\includegraphics[width=.97\linewidth]{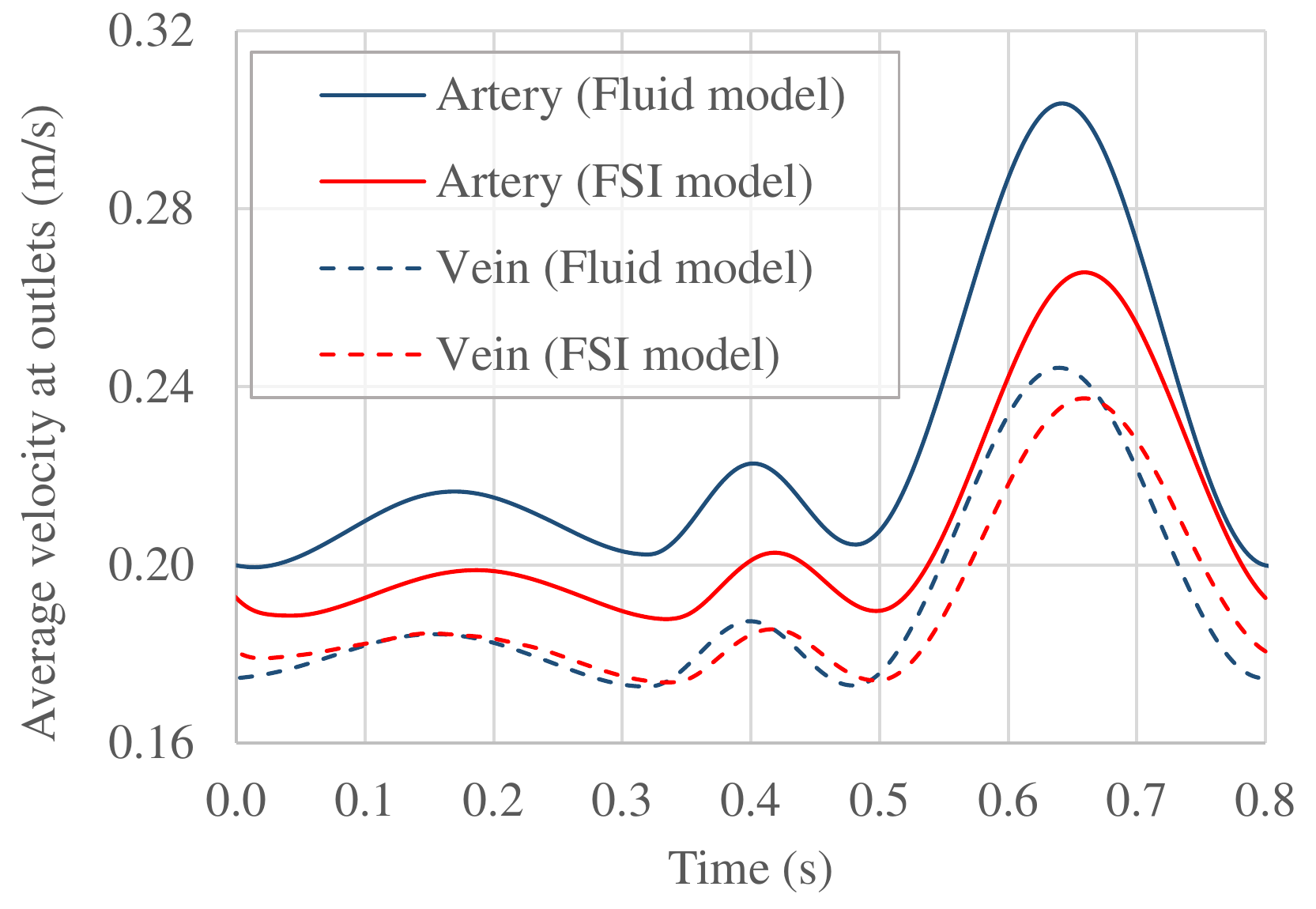}
			\hspace*{1mm}
			\caption{}\label{fig:OutletVelocitiesComp_Pub} 
		\end{subfigure}
		\begin{subfigure}{.49\textwidth}
			\centering
			\hspace*{0.1mm}
			\includegraphics[width=.97\linewidth]{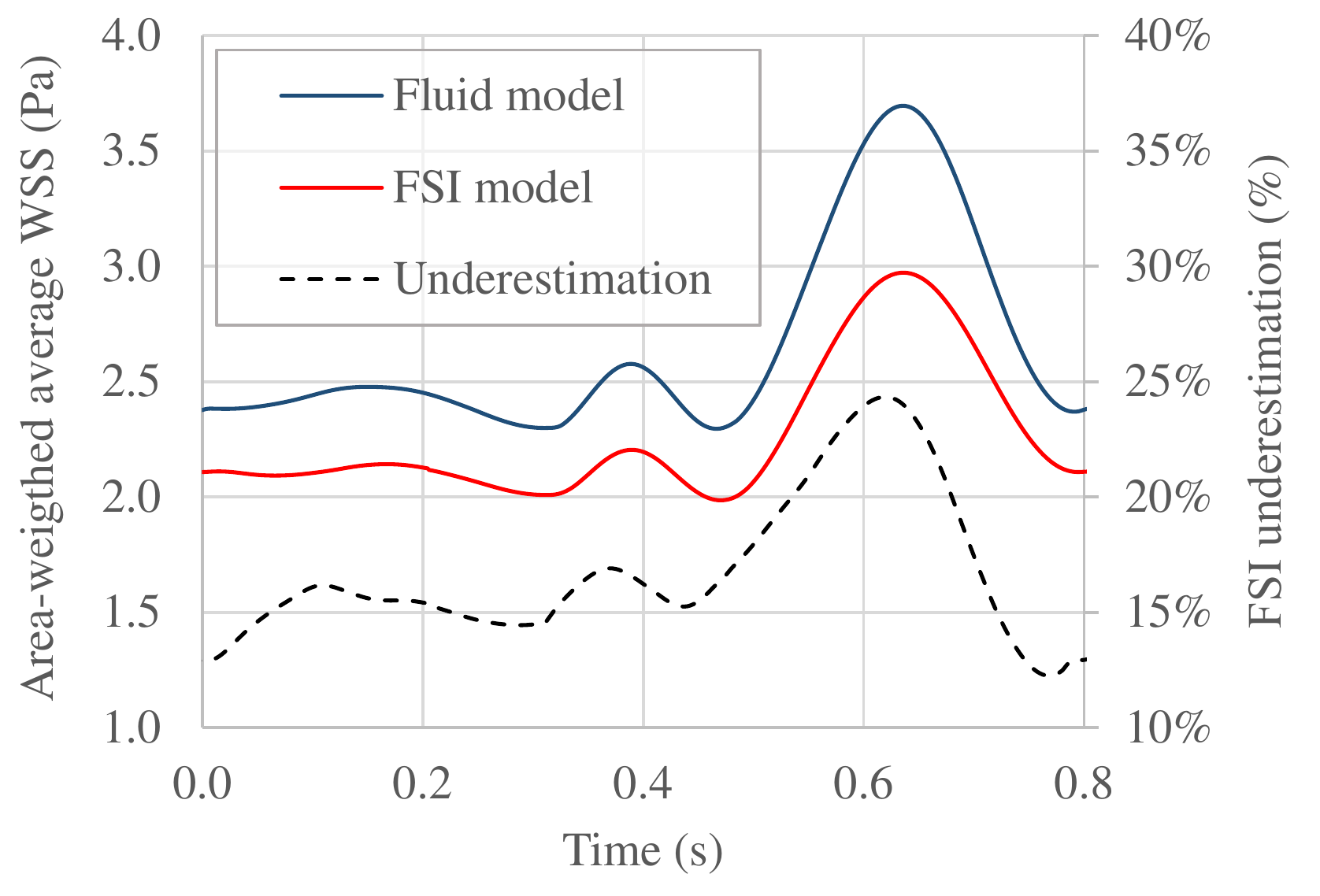}
			\caption{}\label{fig:AverageWSSComp_Pub} 
		\end{subfigure}
		\caption[Maximum deformation and principal strain profiles in the vessel walls]{Comparison of the cardiac-cycle waveforms between the fluid and FSI models: \subref{fig:OutletVelocitiesComp_Pub} average outlet velocities and \subref{fig:AverageWSSComp_Pub} average WSS over the vessel walls.} \label{fig:AverageWSSVelComp_Pub}
	\end{figure}

	The reduction in velocity of the blood flow also leads to a large underestimation of the WSSs by the FSI model particularly at the anastomosis, this is evident in Figure~\ref{fig:WSSComp}. In Figures~\ref{fig:OutletVelocitiesComp_Pub} and \ref{fig:AverageWSSComp_Pub} one can see that the average velocities and WSSs are lower throughout the cardiac-cycle for the FSI model. If the effects of prestressing are disregarded, and the FSI model solution is taken as the baseline solution, the peak WSS in the fluid model is found to be overestimated by \SI{12}{\percent}. This is close to the corresponding overestimation of \SI{15}{\percent} stated in the work by Decorato et al. where prestressing was not taken into account \citep{decorato2014numerical}.
		
\begin{figure}
		\begin{subfigure}{.545\textwidth}
			\centering
			\includegraphics[width=.8\textwidth]{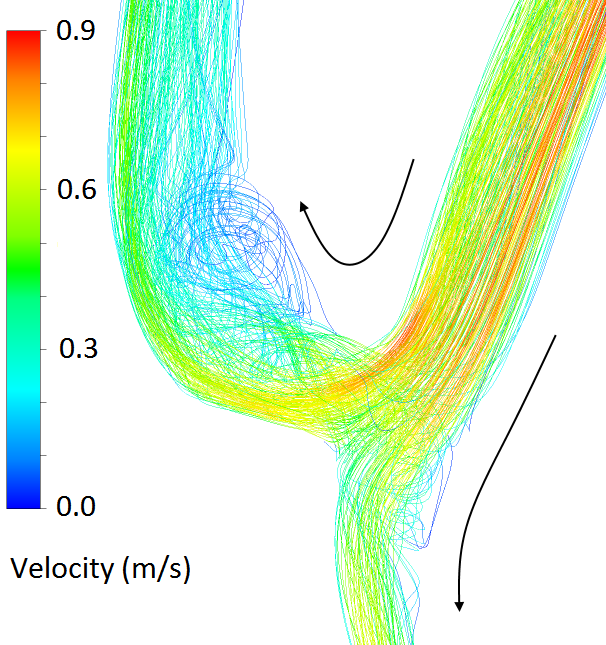}
			\caption{} \label{fig:mri_vel_comp1}
		\end{subfigure}
		\begin{subfigure}{.455\textwidth}
			\centering
			\includegraphics[width=.8\textwidth]{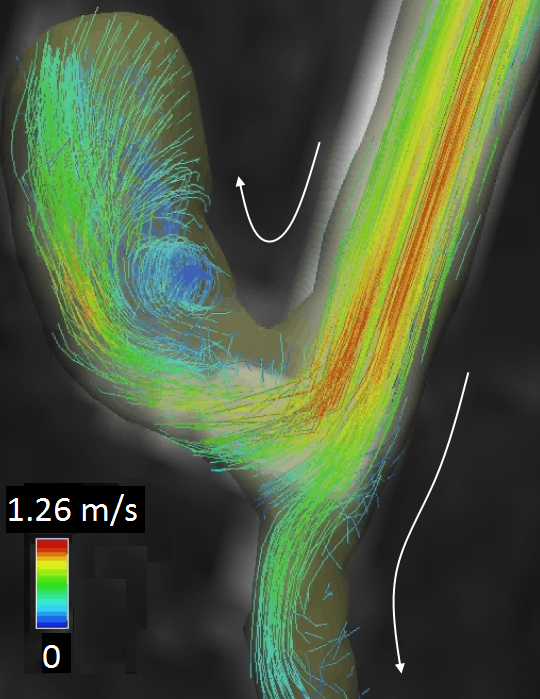}
			\caption{} \label{fig:fluid_vel_comp1}
		\end{subfigure}			
		\caption[PC-MRA and fluid model velocity streamlines]{Qualitative comparison of velocity streamlines: \subref{fig:mri_vel_comp1} PC-MRA sequence and processing \protect\citep{Jermy} and \subref{fig:fluid_vel_comp1} fluid model solution} \label{fig:vel_comp2}
	\end{figure}

	The velocity streamlines of the fluid model are shown alongside the PC-MRA velocity streamlines in Figure~\ref{fig:vel_comp2}. These results are at taken at the time of peak flow rate. The fistula is viewed such that the artery is on the right with blood flow approaching the junction from the top right  of the image. Since the flow profile imposed at the inlet is not the patient-specific profile (see Section~\ref{sec:MRA}) the velocity magnitudes cannot really be compared here. It is evident however that the fluid model solution predicts the same flow features as the MRI sequence and that the relative velocity distributions are also comparable.
	
	These results show that the FSI and fluid models predict very similar flow patterns despite the large difference in the geometry of the models due to neglecting to prestress the FSI model. Both models also agree qualitatively with the PC-MRA data in terms of flow features despite the fact that the flow conditions are not that of the scanned patient. The WSS are the important criteria in the haemodynamics of vascular access, however, and since the WSS are dependant on the velocities, an accurate solution to the velocity distribution is crucial. Prestressing is therefore vital to accurately predict the velocities and WSSs in FSI models of cardiovascular vessels developed from MRI data. The comparative error between a prestressed FSI model, and a fluid model, of a patient-specific fistula therefore still needs to be deduced.
\section{Conclusions and recommendations}
The cardiac-cycle solution of a partitioned FSI model of a patient-specific fistula is presented. The approach has made use of ANSYS Fluent and ANSYS Mechanical software, in conjunction with a semi-implicit staggered coupling algorithm, implemented in ANSYS Workbench software. The fluid is solved iteratively, providing the principal stabilisation features to the FSI model, while the solid direct solver converges fully and provides no form of stabilisation to the staggered scheme. The FSI model is optimally stabilised to minimise its computational expense by combining a number of solution damping techniques within the iterative fluid solver, including FSI boundary stabilisation, together with a boundary transfer ramping applied to the coupling algorithm. This validated model is available via Zenodo as a basis for further investigation.

A spring stiffness-based Robin boundary condition successfully mimics the effect of the surrounding tissue on the vessel walls. By combining this method with frictionless boundary conditions at the outlets of the model, it was possible to reduce the problem size and computational expense without overconstraining it. The artery and vein walls have been modelled as an incompressible isotropic hyperelastic material model each with their characteristic strengths. This material model could be enhanced to include anisotropy. The vessel walls are modelled with a varying thickness at the anastomosis, using one layer of enhanced assumed strain solid elements to optimise the efficiency of the FSI model while solving the displacement solution with sufficient accuracy. A three-element Windkessel model was successfully implemented at the outlets of the blood flow domain to simulate the resistance of the upstream and downstream vessels and produce flow and pressure responses found in branching artery and vein structures.

PC-MRA data was used to develop patient-specific models that mimic the flow conditions imaged in an arteriovenous fistula. The combination of MRI and computational modelling is critical to obtaining a better understanding of the haemodynamics in vascular access. MRI imaging can provide velocity data throughout the vascular access from which kinetics of the blood can be inferred while the boundary data can be used in computational modelling to provide additional insight at a far higher resolution.

The soft tissue model deforms excessively if it is not prestressed, this results in an underapproximation of the velocity and WSS solutions, particularly in the artery. In this case, a fluid model that disregards the movement of the vessel walls is likely to be a better predictor of the flow field than an FSI model. Regardless, it has been shown that both models are able to predict the same flow features, including a region of recirculation that could stimulate neointimal hyperplasia and lead to vascular access failure.

Models such as the one developed in this work may provide valuable data for assessment and optimisation of fistula and grafts for current and future haemodialysis patients. Adjustments and improvements to these types of models may advance our understanding of the haemodynamics in vascular access and improve maturation patency, long-term patency and interventional strategies.

\acks
W.G., B.D.R., and A.M. acknowledge funding from the National Research Foundation through the South African Research Chair in Computational Mechanics (SARCCM, grant agreement SARChI-CM-47584). W.G., J.D., and T.F. acknowledge funding from the National Research Foundation (NRF, grant agreement NRF-CSUR-69007). The views expressed are those of the authors and not necessarily those of the NRF. The authors would like to thank Stephen Jermy, Michael Markl, Ernesta Meintjies, and Delawir Kahn for their work in producing and processing the MRI data. W.G. would like to thank Andie de Villiers for the useful discussions concerning the content of this work. The authors are indebted to Danie de Kock, Daniel Correia, and the rest of the Qfinsoft team for their ANSYS software support and their guidance with modelling in general.

\bibliography{Intro,MRI,references,ANSYS,Books,Results}

\begin{thebibliography}{10}
\providecommand{\url}[1]{\texttt{#1}}
\providecommand{\urlprefix}{URL }
\expandafter\ifx\csname urlstyle\endcsname\relax
  \providecommand{\doi}[1]{doi:\discretionary{}{}{}#1}\else
  \providecommand{\doi}{doi:\discretionary{}{}{}\begingroup
  \urlstyle{rm}\Url}\fi

\bibitem{Vazquez2009}
Vazquez MA. Vascular access for dialysis: recent lessons and new insights.
  \emph{Current Opinion in Nephrology and Hypertension}  2009;
  \textbf{18}(2):116--21.

\bibitem{Liyanage2015}
Liyanage T, Ninomiya T, Jha V, Neal B, Patrice HM, Okpechi I, Zhao M, Lv J.
  Worldwide access to treatment for end-stage kidney disease: a systematic
  review. \emph{The Lancet}  2015; \textbf{385}(9981):1975--1982.

\bibitem{FMC2014}
{ESRD Patients in 2013: A Global Perspective}. \emph{Technical {R}eport},
  Fresenuis Medical Care, Bad Homburg 2014.

\bibitem{schwab1999vascular}
Schwab SJ. Vascular access for hemodialysis. \emph{Kidney International}  1999;
  \textbf{55}(5):2078--2090.

\bibitem{kdoqi2006}
Clinical practice guidelines for vascular access. \emph{American Journal of
  Kidney Diseases}  2006; \textbf{48, Supplement 1}:S248 -- S273.

\bibitem{Nahas2005}
El-Nahas M. The global challenge of chronic kidney disease. \emph{{Kidney
  International}}  2005; \textbf{68}:2918--2929.

\bibitem{Barsoum2006}
Barsoum R. Chronic kidney disease in the developing world. \emph{{The New
  England Journal of Medicine}}  2006; \textbf{354}:997--999.

\bibitem{jain2012global}
Jain AK, Blake P, Cordy P, Garg AX. Global trends in rates of peritoneal
  dialysis. \emph{Journal of the American Society of Nephrology}  2012;
  \textbf{23}(3):533--544.

\bibitem{mcgee2003preventing}
McGee DC, Gould MK. Preventing complications of central venous catheterization.
  \emph{New England Journal of Medicine}  2003; \textbf{348}(12):1123--1133.

\bibitem{windus1993permanent}
Windus DW. Permanent vascular access: a nephrologist's view. \emph{American
  Journal of Kidney Diseases}  1993; \textbf{21}(5):457--471.

\bibitem{schwab1997nkf}
Schwab S, Besarab A, Beathard G, Brouwer D, Etheredge E, Hartigan M, Levine M,
  McCann R, Sherman R, Trerotola S. {NKF-DOQI} clinical practice guidelines for
  vascular access. \emph{American Journal of Kidney Diseases}  1997;
  \textbf{30}(4):S150--S191.

\bibitem{hazinedaroglu2004immediate}
Hazinedaroglu SM, Kayaoglu HA, Ayli D, Duman N, Yerdel MA. Immediate
  postimplant hemodialysis through a new self-sealing heparin-bonded
  polycarbonate/urethane graft. \emph{Transplantation Proceedings}  2004;
  \textbf{36}(9):2599 -- 2602.

\bibitem{Kharboutly2007}
Kharboutly Z, Fenech M, Treutenaere JM, Claude I, Legallais C. Investigations
  into the relationship between hemodynamics and vascular alterations in an
  established arteriovenous fistula. \emph{Medical Engineering \& Physics}
  2007; \textbf{29}(9):999--1007.

\bibitem{Kumwenda2015}
Kumwenda M, Mitra S, Reid C. {Clinical Practice Guidline: Vascular Access For
  Haemodialysis}. \emph{Technical {R}eport}, UK Renal Association 2015.

\bibitem{wasse2007association}
Wasse H, Kutner N, Zhang R, Huang Y. Association of initial hemodialysis
  vascular access with patient-reported health status and quality of life.
  \emph{Clinical Journal of the American Society of Nephrology}  2007;
  \textbf{2}(4):708--714.

\bibitem{allon2007current}
Allon M. Current management of vascular access. \emph{Clinical Journal of the
  American Society of Nephrology}  2007; \textbf{2}(4):786--800.

\bibitem{ravani2007clinical}
Ravani P, Spergel LM, Asif A, Roy-Chaudhury P, Besarab A. Clinical epidemiology
  of arteriovenous fistula in 2007. \emph{Journal of Nephrology}  2007;
  \textbf{20}(B):141.

\bibitem{fan1992vascular}
Fan PY, Schwab SJ. Vascular access: concepts for the 1990s. \emph{Journal of
  the American Society of Nephrology}  1992; \textbf{3}(1):1--11.

\bibitem{harland1994placement}
Harland RC. Placement of permanent vascular access devices: Surgical
  considerations. \emph{Advances in Renal Replacement Therapy}  1994;
  \textbf{1}(2):99--106.

\bibitem{munda1983}
Munda R, First MR, Alexander JW, Linnemann CC, Fidler JP, Kittur D.
  Polytetrafluoroethylene graft survival in hemodialysis. \emph{The Journal of
  the American Medical Association}  1983; \textbf{249}(2):219--222.

\bibitem{lee2015new}
Lee T, Haq NU. New developments in our understanding of neointimal hyperplasia.
  \emph{Advances in Chronic Kidney Disease}  2015; \textbf{22}(6):431--437.

\bibitem{safa1996detection}
Safa AA, Valji K, Roberts AC, Ziegler TW, Hye RJ, Oglevie SB. Detection and
  treatment of dysfunctional hemodialysis access grafts: effect of a
  surveillance program on graft patency and the incidence of thrombosis.
  \emph{Radiology}  1996; \textbf{199}(3):653--657.

\bibitem{roy2001venous}
Roy-Chaudhury P, Kelly BS, Miller MA, Reaves A, Armstrong J, Nanayakkara N,
  Heffelfinger SC. Venous neointimal hyperplasia in polytetrafluoroethylene
  dialysis grafts. \emph{Kidney International}  2001;
  \textbf{59}(6):2325--2334.

\bibitem{lee2009advances}
Lee T, Roy-Chaudhury P. Advances and new frontiers in the pathophysiology of
  venous neointimal hyperplasia and dialysis access stenosis. \emph{Advances in
  Chronic Kidney Disease}  2009; \textbf{16}(5):329--338.

\bibitem{EneIordache2001}
Ene-Iordache B, Mosconi L, Remuzzi G, Remuzzi A. Computational fluid dynamics
  of a vascular access case for hemodialysis. \emph{Journal of Biomechanical
  Engineering}  2001; \textbf{123}(3):284, \doi{10.1115/1.1372702}.

\bibitem{niemann2011computational}
Niemann AK, Thrysoe S, Nygaard JV, Hasenkam JM, Petersen SE. Computational
  fluid dynamics simulation of av fistulas: from {MRI} and ultrasound scans to
  numeric evaluation of hemodynamics. \emph{The Journal of Vascular Access}
  2011; \textbf{13}(1):36--44.

\bibitem{chen2009effect}
Chen J, Wang S, Ding G, Yang X, Li H. The effect of aneurismal-wall mechanical
  properties on patient-specific hemodynamic simulations: two clinical case
  reports. \emph{Acta Mechanica Sinica}  2009; \textbf{25}(5):677--688.

\bibitem{torii2009fluid}
Torii R, Wood NB, Hadjiloizou N, Dowsey AW, Wright AR, Hughes AD, Davies J,
  Francis DP, Mayet J, Yang GZ, \emph{et~al.}. Fluid--structure interaction
  analysis of a patient-specific right coronary artery with physiological
  velocity and pressure waveforms. \emph{Communications in Numerical Methods in
  Engineering}  2009; \textbf{25}(5):565--580.

\bibitem{xiong2011simulation}
Xiong G, Figueroa CA, Xiao N, Taylor CA. Simulation of blood flow in deformable
  vessels using subject-specific geometry and spatially varying wall
  properties. \emph{International Journal for Numerical Methods in Biomedical
  Engineering}  2011; \textbf{27}(7):1000--1016.

\bibitem{decorato2014numerical}
Decorato I, Kharboutly Z, Vassallo T, Penrose J, Legallais C, Salsac AV.
  Numerical simulation of the fluid structure interactions in a compliant
  patient-specific arteriovenous fistula. \emph{International Journal for
  Numerical Methods in Biomedical Engineering}  2014; \textbf{30}(2):143--159.

\bibitem{markl2003time}
Markl M, Chan FP, Alley MT, Wedding KL, Draney MT, Elkins CJ, Parker DW, Wicker
  R, Taylor CA, Herfkens RJ, \emph{et~al.}. Time-resolved three-dimensional
  phase-contrast {MRI}. \emph{Journal of Magnetic Resonance Imaging}  2003;
  \textbf{17}(4):499--506.

\bibitem{grotenhuis2009validation}
Grotenhuis HB, Westenberg JJ, Steendijk P, van~der Geest RJ, Ottenkamp J, Bax
  JJ, Jukema JW, de~Roos A. Validation and reproducibility of aortic pulse wave
  velocity as assessed with velocity-encoded {MRI}. \emph{Journal of Magnetic
  Resonance Imaging}  2009; \textbf{30}(3):521--526.

\bibitem{harloff20093d}
Harloff A, Albrecht F, Spreer J, Stalder A, Bock J, Frydrychowicz A,
  Sch{\"o}llhorn J, Hetzel A, Schumacher M, Hennig J, \emph{et~al.}. {3D blood
  flow characteristics in the carotid artery bifurcation assessed by
  flow-sensitive 4D MRI at 3T}. \emph{Magnetic Resonance in Medicine}  2009;
  \textbf{61}(1):65--74.

\bibitem{frydrychowicz2009three}
Frydrychowicz A, Stalder AF, Russe MF, Bock J, Bauer S, Harloff A, Berger A,
  Langer M, Hennig J, Markl M. Three-dimensional analysis of segmental wall
  shear stress in the aorta by flow-sensitive four-dimensional-{MRI}.
  \emph{Journal of Magnetic Resonance Imaging}  2009; \textbf{30}(1):77--84.

\bibitem{gharahi2016computational}
Gharahi H, Zambrano BA, Zhu DC, DeMarco JK, Baek S. Computational fluid dynamic
  simulation of human carotid artery bifurcation based on anatomy and
  volumetric blood flow rate measured with magnetic resonance imaging.
  \emph{International Journal of Advances in Engineering Sciences and Applied
  Mathematics}  2016; \textbf{8}(1):46--60.

\bibitem{canstein20083d}
Canstein C, Cachot P, Faust A, Stalder A, Bock J, Frydrychowicz A, K{\"u}ffer
  J, Hennig J, Markl M. {3D MR} flow analysis in realistic rapid-prototyping
  model systems of the thoracic aorta: Comparison with in vivo data and
  computational fluid dynamics in identical vessel geometries. \emph{Magnetic
  Resonance in Medicine}  2008; \textbf{59}(3):535--546.

\bibitem{guess2017model}
Guess W. {ANSYS Workbench Release 17 project: Fluid-structure interaction model
  of a patient-specific arteriovenous access fistula} 2017,
  \doi{10.5281/zenodo.376996}.
  \urlprefix\url{http://dx.doi.org/10.5281/zenodo.376996}.

\bibitem{Galpin1995}
Galpin PF, Broberg RB, Hutchinson BR. Three-dimensional navier stokes
  predictions of steady-state rotor/stator interaction with pitch change.
  \emph{Proceedings of 3rd Annual Conference of the CFD Society of Canada,
  Banff, AB, Canada}, vol. 3rd Annual Conference of the CFD, Advanced
  Scientific Computing Ltd: Society of Canada, Banff, Alberta, Canada, 1995.

\bibitem{jansen1992fast}
Jansen KE, Shakib F, Hughes TJR. Fast projection algorithm for unstructured
  meshes. \emph{Computational Nonlinear Mechanics in Aerospace Engineering}
  1992; \textbf{146}:175.

\bibitem{ngoepe2011numerical}
Ngoepe MN, Reddy BD, Kahn D, Meyer C, Zilla P, Franz T. A numerical tool for
  the coupled mechanical assessment of anastomoses of {PTFE} arterio-venous
  access grafts. \emph{Cardiovascular Engineering and Technology}  2011;
  \textbf{2}(3):160--172.

\bibitem{Degroote2008}
Degroote J, Bruggeman P, Haelterman R, Vierendeels J. Stability of a coupling
  technique for partitioned solvers in {FSI} applications. \emph{Computers \&
  Structures}  2008; \textbf{86}(23-24):2224--2234.

\bibitem{forster2007artificial}
F{\"o}rster C, Wall WA, Ramm E. Artificial added mass instabilities in
  sequential staggered coupling of nonlinear structures and incompressible
  viscous flows. \emph{Computer Methods in Applied Mechanics and Engineering}
  2007; \textbf{196}(7):1278--1293.

\bibitem{ANSYSFluentUsers}
\emph{ANSYS\textsuperscript{\textregistered} Academic Research, Release 16.0,
  Help System, Fluent User's Guide}. ANSYS, Inc., 2015.

\bibitem{Jermy}
Jermy S. {4D Flow and Displacement Sensitive MR Imaging of Upper-Arm
  Arteriovenous Connections for Haemodialysis}. Master's {T}hesis, University
  of Cape Town 2016. \urlprefix\url{url =
  {http://open.uct.ac.za/handle/11427/20492},}.

\bibitem{markl20124d}
Markl M, Frydrychowicz A, Kozerke S, Hope M, Wieben O. {4D flow MRI}.
  \emph{Journal of Magnetic Resonance Imaging}  2012;
  \textbf{36}(5):1015--1036.

\bibitem{essabbah1981transient}
Essabbah H, Lacombe C, Fabre G, Saint-Blancard J, Daveloose D, Molle D,
  Leterrier F. Transient rheological study of blood stored in a liquid state.
  \emph{Revue Francaise de Transfusion et Immuno-Hematologie}  1981;
  \textbf{24}(4):357--373.

\bibitem{shibeshi2005rheology}
Shibeshi SS, Collins WE. The rheology of blood flow in a branched arterial
  system. \emph{Applied Rheology}  2005; \textbf{15}(6):398.

\bibitem{boyd2007analysis}
Boyd J, Buick JM, Green S. {Analysis of the Casson and Carreau-Yasuda
  non-Newtonian blood models in steady and oscillatory flows using the lattice
  Boltzmann method}. \emph{Physics of Fluids (1994-present)}  2007;
  \textbf{19}(9):093\,103.

\bibitem{hale1955velocity}
Hale JF, McDonald DA, Womersley JR. {Velocity profiles of oscillating arterial
  flow, with some calculations of viscous drag and the Reynolds number}.
  \emph{The Journal of Physiology}  1955; \textbf{128}(3):629.

\bibitem{he1994unsteady}
He X, Ku DN. Unsteady entrance flow development in a straight tube.
  \emph{Journal of Biomechanical Engineering}  1994; \textbf{116}(3):355--360.

\bibitem{kim2009coupling}
Kim HJ, Vignon-Clementel IE, Figueroa CA, LaDisa JF, Jansen KE, Feinstein JA,
  Taylor CA. On coupling a lumped parameter heart model and a three-dimensional
  finite element aorta model. \emph{Annals of Biomedical Engineering}  2009;
  \textbf{37}(11):2153--2169.

\bibitem{kim2010patient}
Kim HJ, Vignon-Clementel IE, Coogan JS, Figueroa CA, Jansen KE, Taylor CA.
  Patient-specific modeling of blood flow and pressure in human coronary
  arteries. \emph{Annals of Biomedical Engineering}  2010;
  \textbf{38}(10):3195--3209.

\bibitem{westerhof2009arterial}
Westerhof N, Lankhaar JW, Westerhof BE. The arterial windkessel. \emph{Medical
  \& Biological Engineering \& Computing}  2009; \textbf{47}(2):131--141.

\bibitem{corpataux2002low}
Corpataux JM, Haesler E, Silacci P, Ris HB, Hayoz D. Low-pressure environment
  and remodelling of the forearm vein in {Brescia--Cimino} haemodialysis
  access. \emph{Nephrology Dialysis Transplantation}  2002;
  \textbf{17}(6):1057--1062.

\bibitem{issa1986solution}
Issa RI. Solution of the implicitly discretised fluid flow equations by
  operator-splitting. \emph{Journal of Computational Physics}  1986;
  \textbf{62}(1):40--65.

\bibitem{gutierrez2002automatic}
Gutierrez MA, Pilon PE, Lage SG, Kopel L, Carvalho RT, Furuie SS. Automatic
  measurement of carotid diameter and wall thickness in ultrasound images.
  \emph{Computers in Cardiology}  2002; :359--362.

\bibitem{gasser2006hyperelastic}
Gasser TC, Ogden RW, Holzapfel GA. Hyperelastic modelling of arterial layers
  with distributed collagen fibre orientations. \emph{Journal of the Royal
  Society Interface}  2006; \textbf{3}(6):15--35.

\bibitem{yeoh1993some}
Yeoh OH. Some forms of the strain energy function for rubber. \emph{Rubber
  Chemistry and Technology}  1993; \textbf{66}(5):754--771.

\bibitem{mcgilvray2010biomechanical}
McGilvray KC, Sarkar R, Nguyen K, Puttlitz CM. A biomechanical analysis of
  venous tissue in its normal and post-phlebitic conditions. \emph{Journal of
  Biomechanics}  2010; \textbf{43}(15):2941--2947.

\bibitem{prendergast2003analysis}
Prendergast PJ, Lally C, Daly S, Reid AJ, Lee TC, Quinn D, Dolan F. Analysis of
  prolapse in cardiovascular stents: a constitutive equation for vascular
  tissue and finite-element modelling. \emph{Journal of Biomechanical
  Engineering}  2003; \textbf{125}(5):692--699.

\bibitem{maher2011inelasticity}
Maher E, Creane A, Sultan S, Hynes N, Lally C, Kelly DJ. Inelasticity of human
  carotid atherosclerotic plaque. \emph{Annals of Biomedical Engineering}
  2011; \textbf{39}(9):2445--2455.

\bibitem{ANSYSAPDLTheory}
\emph{ANSYS\textsuperscript{\textregistered} Academic Research, Release 16.0,
  Help System, Mechanical APDL Theory Reference}. ANSYS, Inc., 2015.

\end{thebibliography}
\end{document}